\documentclass[11pt,round]{article}
\usepackage[top=2cm,bottom=2cm, left=2cm, right=2cm]{geometry}
\usepackage{natbib}
\usepackage[colorlinks=true, allcolors=blue]{hyperref}
\usepackage{bm}
\usepackage{graphicx}
\usepackage{natbib}
\usepackage{booktabs}
\usepackage{amsfonts}
\usepackage{amsmath,amssymb,amsthm}
\usepackage{algorithm}
\usepackage{algpseudocode}

\usepackage{caption}
\usepackage{theorem} 
\theoremstyle{break}
\usepackage{mathpazo}
\usepackage{cases} 

\title{Modelling compositional data with structural zero values}
\author{
Omar Alzeley$^1$ and Michail Tsagris$^2$  \\
$^1$ Department of Mathematics, Umm Al--Qura University, Saudi Arabia \\
\href{mailto:oazeley@uqu.edu.sa}{oazeley@uqu.edu.sa} \\
$^2$ Department of Economics \& Institute of Research in Education and Digital Social \\ Sciences and Humanities, University of Crete, Gallos Campus, Rethimno, Greece \\
\href{mailto:mtsagris@uoc.gr}{mtsagris@uoc.gr}}

\begin{document}
\maketitle

\begin{center}
{\bf Abstract}
\end{center}
Compositional data are positive multivariate data whose sum equals 1. A popular method to analyze such data is via log--ratio transformations, which are however not applicable when zero values are present. In this paper we present a conditional logistic normal distribution suitable for compositional data with structural zero values. The model is applicable to arbitrary dimensions and the EM algorithm guarantees a fast implementation. The model accepts predictors and we provide a residual analysis. We further provide principal component analysis and biplots for visualization of compositional data with zero values present. Simulation studies and real data examples illustrate the performance of the model, with and without covariates, and compare it to the Dirichlet analogue.  
\\
\\
\textbf{Keywords}: compositional data, structural zeros, normal distribution

\section{Introduction}
Compositional data are non--negative multivariate data that sum to the same constant, usually taken to be unity for convenience purposes. Their sample space is the standard simplex
\begin{eqnarray*}
\mathbb{S}^d=\left\lbrace(y_1,...,y_D)^T \bigg\vert y_i \geq 0,\sum_{i=1}^Dy_i=1\right\rbrace, 
\end{eqnarray*}
where $D$ denotes the number of variables (better known as components) and $d=D-1$. 

The Dirichlet distribution constitutes the natural candidate for the analysis of such data, as its support coincides with the simplex. \citet{gueorguieva2008} and \citet{hijazi2009} considered Dirichlet regression to accommodate covariates. A well known limitation of Dirichlet distribution, however, is that the off--diagonal elements of the implied covariance matrix are necessarily negative. \citet{ait2003} proposed a more flexible alternative that circumvents this restriction, namely the application of log--ratio transformations, where the data are mapped from the simplex onto Euclidean space, where standard multivariate regression techniques, among others, may be employed. The inverse log--ratio transformation subsequently allows the fitted values to be mapped back onto the simplex.

A further limitation common to both regression frameworks is their inability to accommodate zero values. When zeros are present, the logarithmic transformation is undefined, rendering neither regression model directly applicable. \citet{ait1982} was the first to address this issue, proposing ad hoc zero--replacement strategies \citep{ait2003} with several desirable properties; in particular, certain of these strategies preserve the log--ratio based Euclidean distances between compositional observations. \citet{palarea2008} subsequently proposed more sophisticated model--based approaches than plain zero--value imputation, and \citet{martin2012} developed a robust variant thereof. Implementations of these model--based imputation techniques are available in the \textit{R} package \textit{robCompositions} \citep{templ2026}. Both classes of methods, however, presuppose that the observed zeros arise from rounding or measurement error, and must therefore be applied prior to the regression analysis. Under this assumption, observed zeros are treated as unobserved, small but strictly positive quantities that have been rounded down to zero, for instance, in geological applications, where the detection limit of an instrument may be insufficient to register trace quantities of a chemical element in a rock specimen.

A notable drawback of these imputation techniques is that the presence of even a single zero within a compositional vector necessitates adjustment of all remaining components; this distortion is compounded as the number of zeros in a dataset increases \citep{tsagris2015}. While such adjustment may be warranted when the observed proportions are themselves imprecise, it nonetheless introduces additional variability into the data, and the resulting estimates may prove sensitive to the assumed detection limit \citep{scealy2011}. In the case of structural, or essential, zeros, however, the modification of the observed value is conceptually inappropriate. A representative example arises in economics, where the proportion of household expenditure allocated to tobacco or alcohol may legitimately equal zero.

A second strand of the literature, adopted by \citet{zadora2010,scealy2011,stewart2011}, and, more recently, \citet{tsagris2018}, dispenses with imputation altogether and instead incorporates zero values explicitly. \citet{zadora2010} and \citet{tsagris2018}, for example, model the probability of a zero value separately from the continuous component of the distribution, an approach also outlined by \citet[p.~271--273]{ait2003}. \citet{bear2016} developed Aitchison's conditional logistic normal (CLN) model and derived closed--form estimators, albeit restricted to the three--component case. \citet{scealy2011} proposed a square--root transformation mapping compositional data onto the surface of a hypersphere, under which zero values constitute admissible points, while \citet{leininger2013} implemented spatial regression for compositional data containing numerous zeros by means of a scaling power transformation combined with a latent multivariate normal model.

In the present paper, we develop the CLN model briefly outlined theoretically by \citet[p.~271--273]{ait2003}, extending, unlike \citet{bear2016}, its applicability to arbitrary dimensions and incorporating the option of Euclidean covariates. Estimation is performed via the EM algorithm to obtain maximum likelihood estimates. The CLN model constitutes the analogue of the zero--adjusted Dirichlet (ZAD) model \citep{tsagris2018}, against which it is compared throughout. As with the ZAD model, we emphasise that the CLN model is designed for structural, rather than rounded, zeros. The CLN regression (CLNR) model is similar in spirit to the model proposed \cite{chen2016}, but there are distinct differences. We impute the zero values, implicitly, in order to estimate the parameters and secondly, we estimate the covariance matrix of the residuals, as opposed to \cite{chen2016} who assumed an identity covariance matrix, and third, we focus on cross--sectional and not longitudinal data.

The next section discusses the CLN model, simulation studies illustrate its performance and compare it to ZAD in Section \ref{sec:sims}, while in Section \ref{sec:real} two applications in real data are discussed. Finally, Section \ref{sec:concs} concludes the paper. 

\section{The CLN model}  \label{sec:setup}
The CLN model was first described (in brief) by \citet[~pg. 271--273]{ait2003}, but was never explored. At first we will mention some preliminaries. 

\subsection{The additive log--ratio transformation}
For a composition with all components strictly positive, define the additive log--ratio (alr) transformation as
\begin{equation}   \label{alr}
\bm{z} = \left(\log\frac{y_1}{y_D}, \dots, \log\frac{y_d}{y_D}\right)^\top \in \mathbb{R}^d,
\end{equation}
whose inverse is given by
\begin{equation} \label{alrinv}
\bm{y}=\left(\frac{e^{z_1}}{1+\sum_{j=1}^de^{z_j}}, \dots, \frac{e^{z_d}}{1+\sum_{j=1}^de^{z_j}}, \frac{1}{1+\sum_{j=1}^de^{z_j}}\right)^\top.
\end{equation}
The Jacobian of the alr transformation is equal to $\prod_{j=1}^D y_{j}^{-1}$, and the logistic normal distribution arises when we assume that the alr transformed compositional data follow the multivariate normal in $\mathbb{R}^d$. 

\subsection{The CLN distribution}
\citet[~pg. 271--273]{ait2003} suggested a two part logistic normal where one part contains the zero--free compositional data and the second the compositions with a specific zero values pattern. Following \citet[~pg. 271--273]{ait2003} we assume the latent, fully--observed alr vector follows a multivariate normal, and that some compositions are observed with structural zeros: a subset of components is exactly zero, and only the surviving components (renormalized) are informative. Zeros are treated as exogenous: components that are zero for unit $i$ is fixed, ancillary information, are not generated by the mean/covariance parameters of the normal model (see the ignorability discussion below). Consequently the log--likelihood factors into a piece coming from the composition/pattern indicator and a piece coming from the log--ratio normal model; only the latter enters the EM iterations. We further highlight that our proposal is to extend the zero values composition to multiple zero value patterns. 

A case where the CLN or the CLNR model will fail to fit is when the number of observations with at least one zero values is less than the number of components $D$, because the covariance matrix is not positive definite. 

\subsection{Zero pattern and the projection}
For a row $i$ with $C_i \le D$ nonzero components ($c_i = C_i - 1$), let $\bm{F}(c) = [\bm{I}_c \; -\bm{1}]$ (the $c\times(c+1)$ alr contrast matrix restricted to $c+1$ surviving components) and $\bm{H}(c) = \bm{I}_c + \bm{1}\bm{1}^\top$\footnote{The $\bm{F}$ and $\bm{H}$ matrices were defined by \cite[~pg. 81]{ait2003}}. Let $\bm{S}_m$ be the $C_i \times D$ selection matrix picking out the rows of $\bm{I}_D$ corresponding to nonzero components. Then the fixed $c_i \times d$ matrix
\begin{equation*}
\bm{Q}_i \;=\; \bm{F}(c_i)\, \bm{S}_m \, \bm{F}(d)^\top \bm{H}(d)^{-1}
\end{equation*}
maps the full latent alr vector $\bm{z} \in \mathbb{R}^d$ onto the alr--contrasts of the observed subcomposition,
\begin{equation*}  
\bm{b}_i = alr\left(\text{surviving components of } \bm{y}_i\right) =\bm{Q}_i \bm{z}_i.
\end{equation*}
The $\bm{Q}_i$ depends only on the zero pattern (through $\bm{S}_m$, $c_i$), not on any unknown parameter.

Partition each observation's contribution as $\Pr(\text{pattern}_i)\times f(\bm{b}_i \mid \text{pattern}_i)$. If pattern probabilities $\bm\pi$ are variation-independent of $(\bm{\mu},\bm{\Sigma})$ (or $(\bm{B},\bm{\Sigma})$ with covariates), the full log--likelihood
\begin{equation*}
\ell(\bm\pi,\bm{\mu},\bm{\Sigma}) \;=\; \underbrace{\sum_i \log \Pr(\text{pattern}_i;\bm\pi)}_{\text{depends on } \bm\pi \text{ only}} \;+\; \underbrace{\sum_i \log f(\bm{b}_i \mid \text{pattern}_i;\bm{\mu},\bm{\Sigma})}_{\ell(\bm{\mu},\bm{\Sigma})}
\end{equation*}
separates. With $K$ distinct patterns and pattern counts $n_1,\dots,n_K$ ($\sum_k n_k = n$), the first term is maximized at $\hat\pi_k = n_k/n$, giving the constant
\begin{equation}  \label{eq:pattern-term}
\ell_{\text{pattern}} \;=\; \sum_{k=1}^K n_k \log\!\left(\frac{n_k}{n}\right),
\end{equation}
independent of $(\bm{\mu},\bm{\Sigma})$. Equation~\eqref{eq:pattern-term} plays no role in the EM iterations (it is a constant with respect to $\bm{\mu},\bm{\Sigma},\bm{B}$) but is added once, at convergence, to report a total log--likelihood comparable across models in which the split itself is also treated as estimated (e.g. for AIC/BIC, with one extra estimated parameter per additional pattern beyond the first).

We assume that $R_i$, the pattern indicator recording which components of $\bm{y}_i$ are structurally zero, is generated by a mechanism with parameters $\pi$ that are variation--independent of $(\bm{\mu}, \bm{\Sigma})$, so that the log--likelihood in \ref{eq:pattern-term} separates additively and the two blocks may be maximised in turn. The condition is analogous to the data mechanisms \citep{rubin1976}, though our setting differs in an important respect: $R_i$ does not mask an unobserved true value, since for a structurally zero part the zero is itself the true value. What is required, therefore, is not that the absence of a value be unrelated to its magnitude, but that the process determining which components are structurally zero for the $i$--compositional vector not itself be driven by $(\bm{\mu}, \bm{\Sigma})$, or by $\bm{B}$ in the regression setting. This is a reasonable assumption when zero occurrence reflects a distinct structural or behavioural decision rather than a threshold on the latent compositional propensity $\bm{z}_i$ itself, and we maintain it throughout. 

\subsection{The simple CLN model}  \label{sec:mle}
Let $\bm{z}_i \overset{\text{iid}}{\sim} N_d(\bm{\mu}, \bm{\Sigma}), \ \ i = 1,\dots,n.$
We split the sample into $n_1$ fully--observed rows ($\bm{z}_i$, and denote this set by $\bm{y}_1$) and $n_0 = n - n_1$ rows with zeros (set $\bm{y}_0$), where only $\bm{b}_i = \bm{Q}_i \bm{z}_i$ is observed. The complete--data log--likelihood (dropping the pattern constant) is:
\begin{equation*}
\ell_c(\bm{\mu},\bm{\Sigma}) = -\frac{n}{2}\log\det(2\pi\bm{\Sigma}) - \frac12\sum_{i=1}^n (\bm{z}_i - \bm{\mu})^\top \bm{\Sigma}^{-1} (\bm{z}_i - \bm{\mu}).
\end{equation*}

The steps of the EM algorithm presented below are very similar to the ones implemented for missing value imputation.

\paragraph{E--step:}
For $i \in \bm{y}_1$, $\bm{z}_i$ is known no imputation needed. For $i \in \bm{y}_0$, using the current estimates $(\bm{\mu}^{(t)}, \bm{\Sigma}^{(t)})$, note $\bm{b}_i = \bm{Q}_i\bm{z}_i$ is a linear transform of the jointly normal $\bm{z}_i$, so
\begin{equation*}
\begin{pmatrix}\bm{z}_i \\ \bm{b}_i\end{pmatrix} \sim N\!\left( \begin{pmatrix}\bm{\mu} \\ \bm{Q}_i\bm{\mu}\end{pmatrix},\; \begin{pmatrix} \bm{\Sigma} & \bm{\Sigma} \bm{Q}_i^\top \\ \bm{Q}_i\bm{\Sigma} & \bm{Q}_i\bm{\Sigma}\bm{Q}_i^\top \end{pmatrix} \right).
\end{equation*}
Writing $\bm{S}_{A,i} = \bm{Q}_i\bm{\Sigma}\bm{Q}_i^\top$ and $\bm{K}_i = \bm{\Sigma}\bm{Q}_i^\top \bm{S}_{A,i}^{-1}$, the conditional law of $\bm{z}_i$ given $\bm{b}_i$ is Gaussian with
\begin{subequations}
\begin{align}
E[\bm{z}_i \bm{b}_i] &= \bm{\mu} + \bm{K}_i\left(\bm{b}_i - \bm{Q}_i\bm{\mu}\right) = \hat{\bm{z}}_i, \label{eq:ez} \\ 
Var(\bm{z}_i \bm{b}_i) &= \bm{\Sigma} - \bm{K}_i \bm{Q}_i \bm{\Sigma} = \bm{V}_i. \label{eq:vz}
\end{align}
\end{subequations}
The sufficient statistics needed for the M--step are the conditional expectations of $\bm{z}_i$ and $\bm{z}_i\bm{z}_i^\top$ are given by:
\begin{equation*}
E[\bm{z}_i \bm{b}_i] = \hat{\bm{z}}_i, \ \  E\left[\bm{z}_i\bm{z}_i^\top \bm{b}_i\right] = \bm{V}_i + \hat{\bm{z}}_i\hat{\bm{z}}_i^\top.
\end{equation*}

\paragraph{M--step:}
Maximizing $E[\ell_c \mid \text{data}]$ over $(\bm{\mu},\bm{\Sigma})$ gives the usual normal--model updates with imputed sufficient statistics:
\begin{subequations}
\begin{align}
\bm{\mu}^{(t+1)} &= \frac{1}{n}\left(\sum_{i\in \bm{y}_1} \bm{z}_i + \sum_{i\in \bm{y}_0} \hat{\bm{z}}_i\right), \label{eq:mu-update}\\
\bm{\Sigma}^{(t+1)} &= \frac{1}{n}\left(\sum_{i \in \bm{y}_1} \bm{z}_i\bm{z}_i^\top + \sum_{i\in \bm{y}_0}\left(\bm{V}_i + \hat{\bm{z}}_i\hat{\bm{z}}_i^\top\right)\right) - \bm{\mu}^{(t+1)}\bm{\mu}^{(t+1)\top}. \label{eq:sigma-update}
\end{align}
\end{subequations}
Note the $\bm{V}_i$ term in \eqref{eq:sigma-update}: using only the point imputations $\hat{\bm{z}}_i$ (single imputation) and dropping $\bm{V}_i$ would systematically underestimate $\bm{\Sigma}$, since it ignores the residual uncertainty left in $\bm{z}_i$ after conditioning on the observed subcomposition.

Using the current $\bm{\mu}$ and $\bm{\Sigma}$ estimates we compute the two parts of the log--likelihood,
\begin{subequations}   
\begin{align}
\ell_1(\bm{\mu},\bm{\Sigma}) &= -\frac{n_1}{2}\log\det(2\pi\bm{\Sigma}) - \frac12\sum_{i\in \bm{y}_1}(\bm{z}_i-\bm{\mu})^\top\bm{\Sigma}^{-1}(\bm{z}_i-\bm{\mu}) - \sum_{i \in \bm{y}_1}\sum_{j=1}^D \log y_{ij}, \label{eq:ll1}\\
\ell_0(\bm{\mu},\bm{\Sigma}) &= \sum_{i \in \bm{y}_0}\left[-\frac12\log\det(2\pi \bm{S}_{A,i}) - \frac12(\bm{b}_i - \bm{Q}_i\bm{\mu})^\top \bm{S}_{A,i}^{-1}(\bm{b}_i-\bm{Q}_i\bm{\mu}) - \sum_{j: y_{ij}>0}\log y_{ij} \right]. \label{eq:ll2}
\end{align}
\end{subequations}
For $i\in \bm{y}_0$, $\ell_0$ uses the marginal law of the observed contrast $\bm{b}_i = \bm{Q}_i\bm{z}_i \sim N(\bm{Q}_i\bm{\mu},\, \bm{Q}_i\bm{\Sigma}\bm{Q}_i^\top)$, i.e.\ the density of what was actually observed, integrating out the unobserved directions of $\bm{z}_i$, not the conditional density used in the E--step. The full log--likelihood is 
$\ell(\bm{\mu},\bm{\Sigma}) = \ell_1(\bm{\mu},\bm{\Sigma}) + \ell_0(\bm{\mu},\bm{\Sigma}) + \ell_{\text{pattern}}$. EM iterates \eqref{eq:ez}--\eqref{eq:sigma-update} until $\ell(\bm{\mu},\bm{\Sigma})$ stabilizes.

\subsection{Biplots and PCA}
\cite{ait2002} proposed the biplots for visual display of the compositional data. The EM algorithm above estimates the covariance matrix of the alr transformation. 
Biplots for compositional data though require the covariance matrix computed based on the centered log--ratio (clr) transformation that is defined as
\begin{equation}   \label{clr}
\bm{z}_0 = \left(\log\frac{y_1}{g\left(\bm{y}\right)}, \dots, \log\frac{y_d}{g\left(\bm{y}\right)}\right)^\top,
\end{equation}
where $g\left(\bm{y}\right)=\prod_{j=1}^Dy_j$ is the geometric mean. 
We convert the covariance matrix $\bm{\Sigma}_{alr}$ to the covariance matrix of the clr transformation $\bm{\Sigma}_{clr}$ using Aitchison's relationship \citep[pg.~83]{ait2003}
\begin{equation*}
\bm{\Sigma}_{clr}=\bm{F}(d)^\top\bm{H}(d)^{-1}\bm{\Sigma}_{alr}\bm{H}(d)^{-1}\bm{F}(d).
\end{equation*}

The alr--transformed compositional data with no zeros need no modification, but for the alr--transformed compositional data that contain zero values require attention. For their components that contain zero values, we use the expectations, $\hat{\bm{z}}_i$ values to fill--in those components. These said, we conclude that the constructed biplot is an approximation to the ground--truth biplot that would have been obtained if there were no zero values present.

PCA for compositional data was first proposed by \cite{ait1983}, again using the clr transformation (\ref{clr}). Moving along the same lines as with the biplots we can perform PCA and use it to reduce the dimensionality of the compositional data. 

\subsection{Two--sample equality of mean vector test without assuming equality of the covariance matrices} \label{james}
\citet{james1954} proposed a test for linear form of hypotheses of the population means ($\bm{\mu}_1=\bm{\mu_2}$) when the covariance matrices are unknown and need not be assumed equal. The test statistic for two $d$--dimensional samples is
\begin{eqnarray} \label{T2}
T^2=\left(\hat{\bm {\mu}}_1-\hat{\bm{\mu}}_2\right)^\top\tilde{\bm{\Sigma}}^{-1}\left(\hat{\bm{\mu}}_1-\hat{\bm{\mu}}_2\right), 
\end{eqnarray} 
with $\tilde{\bm{\Sigma}}=\tilde{\bm{\Sigma}}_1+\tilde{\bm{\Sigma}}_2=\frac{\hat{\bm{\Sigma}}_1}{n_1}+\frac{\hat{\bm{\Sigma}}_2}{n_2}$. 
\citet{james1954} suggested that the test statistic is to be compared with $2h\left(\alpha\right)$, a corrected $\chi^2$ distribution whose form is
\begin{eqnarray*}
2h\left(\alpha\right)=\chi^2_{d,1-a}\left(A+B\chi^2_{d,1-a}\right),
\end{eqnarray*} 
where $a$ is the level of significance, $d$ are the degrees of freedom of the $\chi^2$ distribution, and 
\begin{eqnarray*}
A = 1+\frac{1}{2d}\sum_{i=1}^2\frac{\left(tr \tilde{\bm{\Sigma}}^{-1}\tilde{\bm{\Sigma}_i}\right)^2}{n_i-1} \ \ \text{and} \ \ 
B = \frac{1}{d\left(d+2\right)}\left[\frac{1}{2}\sum_{i=1}^2\frac{tr\left(\tilde{\bm{\Sigma}}^{-1}\tilde{\bm{\Sigma}}_i\right)^2}{n_i-1}
+\frac{1}{2}\sum_{i=1}^2\frac{\left(tr \tilde{\bm{\Sigma}}^{-1}\tilde{\bm{\Sigma}}_i\right)^2}{n_i-1} \right].
\end{eqnarray*}

Since the test statistic (\ref{T2}) requires only the estimated mean vectors and  covariance matrices, we may fit the CLN model to each of the two groups and plug in the estimated parameters.  

\subsection{The CLNR model} \label{sec:reg}
Let $\bm{x}_i \in \mathbb{R}^p$ be fully observed covariates (including an intercept). Replace the constant mean by a linear predictor:
\begin{equation*}
\bm{z}_i  \bm{x}_i \sim N_d\left(\bm{B}^\top \bm{x}_i, \ \  \bm{\Sigma}\right), \ \ \bm{B} \in \mathbb{R}^{p\times d}.
\end{equation*}
Everything about the zero pattern and $\bm{Q}_i$ from Section~\ref{sec:setup} carries over unchanged $\bm{Q}_i$ is purely geometric (depends on which components are zero), not on the mean model.

\paragraph{E--step:}
Now the conditional mean of $\bm{z}_i$ is row--specific, $\bm{\mu}_i = \bm{B}^\top \bm{x}_i$, but the conditioning algebra is otherwise identical to \eqref{eq:ez}--\eqref{eq:vz}:
\begin{subequations}   
\begin{align} 
\hat{\bm{z}}_i &= \bm{\mu}_i + \bm{K}_i\left(\bm{b}_i - \bm{Q}_i \bm{\mu}_i\right) = \bm{B}^\top\bm{x}_i + \bm{K}_i\left(\bm{b}_i - \bm{Q}_i \bm{B}^\top\bm{x}_i\right), \label{eq:ez-reg} \\
\bm{V}_i &= \bm{\Sigma} - \bm{K}_i\bm{Q}_i\bm{\Sigma}. \label{eq:vz-reg}
\end{align}
\end{subequations}
Note $\bm{V}_i$ does not depend on $\bm{x}_i$ or $\bm{B}$ (it depends only on $\bm{\Sigma}$ and the pattern--determined $\bm{Q}_i$), only the conditional mean $\hat{\bm{z}}_i$ picks up covariate information, where
\begin{equation*}
\bm{S}_{A,i} = \bm{Q}_i \bm{\Sigma} \bm{Q}_i^\top, \qquad \bm{K}_i = \bm{\Sigma}\bm{Q}_i^\top \bm{S}_{A,i}^{-1}.
\end{equation*}

\paragraph{M--step:}
Stack the completed responses $\bm{Z} = \begin{pmatrix}\bm{Z}_{\text{full}}\\ \hat{\bm{Z}}\end{pmatrix}$ ($\bm{z}_{full,i}$, $i\in \bm{y}_1$, and $\hat{\bm{z}}_i$, $i \in \bm{y}_0$) and covariates $\bm{X} = \begin{pmatrix}\bm{x}_1^\top;\dots \bm{x}_p^\top\end{pmatrix}$ correspondingly. Maximizing $E[\ell_c \mid \text{data}]$ over $\bm{B}$ for fixed $\bm{\Sigma}$ gives a generalized least-squares problem
\begin{equation*}
\max_{\bm{B}} \; -\frac12 \sum_i \left(\bm{z}_i - \bm{B}^\top\bm{x}_i\right)^\top \bm{\Sigma}^{-1}\big(\bm{z}_i - \bm{B}^\top\bm{x}_i\big)
\end{equation*}
with a common $\bm{\Sigma}^{-1}$ weight across all rows; the weight cancels out of the normal equations, so GLS reduces to plain OLS:
\begin{equation}
\bm{B}^{(t+1)} = \big(\bm{X}^\top \bm{X}\big)^{-1}\bm{X}^\top \bm{Z}. \label{eq:B-update}
\end{equation}
Given $\bm{B}^{(t+1)}$, residuals $\bm{r}_i = \bm{z}_i - \bm{B}^{(t+1)\top}\bm{x}_i$ (using $\bm{z}_i$ for $\bm{y}_1$, $\hat{\bm{z}}_i$ for $\bm{y}_0$), and
\begin{equation}
\bm{\Sigma}^{(t+1)} = \frac{1}{n}\left(\sum_{i \in \bm{y}_1} \bm{r}_i\bm{r}_i^\top + \sum_{i\in \bm{y}_0}\big(\bm{r}_i\bm{r}_i^\top + \bm{V}_i\big)\right). \label{eq:Sigma-update-reg}
\end{equation}
As before, $\sum_{i\in \bm{y}_0}\bm{V}_i$ is the imputation--uncertainty correction; omitting it biases $\hat{\bm{\Sigma}}$ downward.
\begin{subequations}   
\begin{align}
\ell_1(\bm{B},\bm{\Sigma}) &= -\frac{n_1}{2}\log\det(2\pi\bm{\Sigma}) - \frac12 \sum_{i \in \bm{y}_1} \big(\bm{z}_i - \bm{B}^\top\bm{x}_i\big)^\top \bm{\Sigma}^{-1}\big(\bm{z}_i - \bm{B}^\top\bm{x}_i\big) - \sum_{i\in \bm{y}_1}\sum_{j=1}^D \log y_{ij}, \label{eq:ll1-reg}\\
\ell_0(\bm{B},\bm{\Sigma}) &= \sum_{i\in \bm{y}_0}\Bigg[-\frac12\log\det(2\pi\bm{S}_{A,i}) - \frac12\big(\bm{b}_i - \bm{Q}_i\bm{B}^\top\bm{x}_i\big)^\top \bm{S}_{A,i}^{-1}\big(\bm{b}_i - \bm{Q}_i\bm{B}^\top\bm{x}_i\big) - \sum_{j: y_{ij}>0}\log y_{ij}\Bigg]. \label{eq:ll2-reg}
\end{align}
\end{subequations}
Again $\bm{S}_{A,i} = \bm{Q}_i\bm{\Sigma}\bm{Q}_i^\top$ uses the marginal covariance of the observed contrast, and the pattern term \eqref{eq:pattern-term}, computed here from the fully--positive/has--zeros split, or more generally from covariate--stratified pattern frequencies if $\bm{\pi}$ is allowed to depend on $\bm{x}_i$ is a separate, additive, non--iterated constant:
\begin{equation*}
\ell(\bm{B},\bm{\Sigma}) = \ell_1(\bm{B},\bm{\Sigma}) + \ell_0(\bm{B},\bm{\Sigma}) +\ell_{\text{pattern}}.
\end{equation*}

Setting $\bm{x}_i \equiv 1$ (intercept--only design, $p=1$) collapses the regression model to that of Section~\ref{sec:mle}: $\bm{B}$ becomes a $1\times d$ row vector equal to $\bm{\mu}^\top$, \eqref{eq:B-update} reduces to $\bm{\mu} = \frac{1}{n}\sum_i \hat{\bm{z}}_i$ matching \eqref{eq:mu-update}, and \eqref{eq:Sigma-update-reg} matches \eqref{eq:sigma-update} once $\bm{\mu}\bm{\mu}^\top$ is expanded from the residual sum of squares. 

\subsubsection{Starting values and online updating}
The initial values for the CLNR model may be obtained from the multivariate zero--inflated logistic normal regression of \cite{li2018}. However, a computationally more efficient method is to compute the ordinary least squares estimates using the fully observed compositional data and their corresponding predictor variable values. When new observations are available, one should update the regression coefficients by performing the EM algorithm in the integrated data. In this case, to reduce the number of iterations, the starting values will be the previously estimated regression coefficients. 

\subsubsection{Covariance matrix of the regression coefficients}
The EM algorithm above targets the observed--data log--likelihood $\ell(\bm{B},\bm{\Sigma})=\ell_1(\bm{B},\bm{\Sigma})+\ell_0(\bm{B},\bm{\Sigma})$ from \eqref{eq:ll1-reg}--\eqref{eq:ll2-reg}, but it never differentiates $\ell$ directly: for $i\in\bm{y}_0$, $\ell_0$ involves only the marginal law of the observed contrast $\bm{b}_i$, whose curvature with respect to $(\bm{B},\bm{\Sigma})$ is considerably harder to work with than that of the complete--data log--likelihood
\begin{equation}
\ell_c(\bm{B},\bm{\Sigma}) = -\frac{n}{2}\log\det(2\pi\bm{\Sigma}) - \frac12 \sum_{i=1}^n \big(\bm{z}_i - \bm{B}^\top\bm{x}_i\big)^\top \bm{\Sigma}^{-1} \big(\bm{z}_i - \bm{B}^\top\bm{x}_i\big),
\end{equation}
the same $\ell_c$ that underlies the E-- and M--steps in Section~\ref{sec:reg}. Louis' identity \citep{louis1982} exploits exactly this: it recovers the curvature of the harder, observed--data $\ell$ using only quantities built from the easier, complete--data $\ell_c$, conditioned on what was actually observed.

Write the parameter as $\bm\theta = \big(\text{vec}(\bm{B}), \text{vech}(\bm{\Sigma})\big)$, where $\text{vech}(\bm{\Sigma})$ stacks the $d(d+1)/2$ distinct entries of the symmetric matrix $\bm{\Sigma}$ (its lower triangle, including the diagonal), while $\text{vec}(\bm{\Sigma})$ stacks all $d^2$ entries and so repeats each off--diagonal entry twice; the two are related by the duplication matrix $\bm{G}_d$, the fixed $d^2\times\tfrac{d(d+1)}{2}$ zero--one matrix satisfying $\text{vec}(\bm{\Sigma}) = \bm{G}_d\,\text{vech}(\bm{\Sigma})$ for every symmetric $\bm{\Sigma}$; it simply routes each entry of $\text{vech}(\bm{\Sigma})$ to the one or two positions it occupies in $\text{vec}(\bm{\Sigma})$. Define the complete--data score and information as the gradient and negative Hessian of $\ell_c$ with respect to $\bm\theta$,
\begin{equation}
\bm{u}_c = \frac{\partial \ell_c(\bm{B},\bm{\Sigma})}{\partial\bm\theta}, \qquad \bm{I}_c = -\frac{\partial^2 \ell_c(\bm{B},\bm{\Sigma})}{\partial\bm\theta\,\partial\bm\theta^\top}.
\end{equation}
Both are ordinary quantities from a standard multivariate normal linear model \emph{if} every $\bm{z}_i$ were observed; the difficulty is that, for $i\in\bm{y}_0$, only $\bm{b}_i=\bm{Q}_i\bm{z}_i$ was actually observed, so $\bm{z}_i$ itself remains uncertain even after the EM algorithm has converged. Louis' identity states that the observed--data information $\bm{I}_{obs}=-\partial^2\ell(\bm{B},\bm{\Sigma})/\partial\bm\theta\partial\bm\theta^\top$, evaluated at the MLE, equals
\begin{equation}   \label{eq:louis}
\bm{I}_{obs} \;=\; E\big[\bm{I}_c \mid \text{data}\big] \;-\; \text{Var}\big[\bm{u}_c \mid \text{data}\big],
\end{equation}
a $\big(pd+\tfrac{d(d+1)}{2}\big)\times\big(pd+\tfrac{d(d+1)}{2}\big)$ matrix. The first term is the complete--data information one would have if every $\bm{z}_i$ were observed, averaged over the remaining uncertainty about the unobserved $\bm{z}_i$, $i\in\bm{y}_0$, given the data; the second term corrects for that same uncertainty by subtracting the conditional variance it induces in the complete--data score $\bm{u}_c$ -- if there were no missing components at all ($\bm{y}_0=\varnothing$), $\bm{u}_c$ would be non--random given the data and \eqref{eq:louis} would reduce to $\bm{I}_{obs}=\bm{I}_c$, the ordinary complete--data information. Both terms of \eqref{eq:louis} turn out to be expressible entirely in terms of quantities the EM algorithm already computes at convergence, so that no new optimization or resampling is required.

For $i \in \bm{y}_1$ write $\bm{r}_i = \bm{z}_i - \hat{\bm{B}}^\top \bm{x}_i$ (known exactly), and for $i \in \bm{y}_0$ write $\hat{\bm{r}}_i = \hat{\bm{z}}_i - \hat{\bm{B}}^\top \bm{x}_i$, the completed residual obtained from the E--step imputation $\hat{\bm{z}}_i$ in \eqref{eq:ez-reg}, evaluated at the converged $(\hat{\bm{B}},\hat{\bm{\Sigma}})$. For $i \in \bm{y}_0$ define the whitened quantities
\begin{equation}
\bm{P}_i = \hat{\bm{\Sigma}}^{-1} \bm{V}_i \hat{\bm{\Sigma}}^{-1}, \qquad \bm{a}_i = \hat{\bm{\Sigma}}^{-1}\hat{\bm{r}}_i,
\end{equation}
where $\bm{V}_i$ is the conditional covariance of $\bm{z}_i$ given $\bm{b}_i$ from \eqref{eq:vz-reg} -- the same quantity that already appears in the M--step update \eqref{eq:Sigma-update-reg}, since it measures exactly how much uncertainty about $\bm{z}_i$ remains after conditioning on the observed subcomposition. Both $\bm{P}_i$ and $\bm{a}_i$ are already available once the E--step has converged.

\paragraph{The $E[\bm{I}_c\mid\text{data}]$ term.} The complete--data information for $\text{vec}(\bm{B})$ is data--independent, $\bm{I}_c^{BB} = \hat{\bm{\Sigma}}^{-1}\otimes \bm{X}^\top \bm{X}$, since it comes from a term in $\ell_c$ that is exactly quadratic in $\bm{B}$; no conditioning on the data is needed for this block. The $\text{vech}(\bm{\Sigma})$ block and the $\big(\text{vec}(\bm{B}),\text{vech}(\bm{\Sigma})\big)$ cross block of $\bm{I}_c$, by contrast, both depend on the residuals $\bm{r}_i$ and hence do require conditioning on the data. Substituting $\bm{r}_i$ ($i\in \bm{y}_1$) and $\hat{\bm{r}}_i$ ($i\in \bm{y}_0$) into the OLS normal equations $\sum_i \bm{x}_i \bm{r}_i^\top = \bm{0}$ underlying \eqref{eq:B-update} and the fixed--point relation \eqref{eq:Sigma-update-reg} defining $\hat{\bm{\Sigma}}$ makes both of these collapse to their familiar unconditional forms at the converged $(\hat{\bm{B}}, \hat{\bm{\Sigma}})$,
\begin{equation}   \label{eq:EIc}
E\big[\bm{I}_c^{\Sigma\Sigma} \mid \text{data}\big] = \frac{n}{2}\,\bm{G}_d^\top\big(\hat{\bm{\Sigma}}^{-1}\otimes\hat{\bm{\Sigma}}^{-1}\big) \bm{G}_d, \qquad E\big[\bm{I}_c^{B\Sigma} \mid \text{data}\big] = \bm{0},
\end{equation}
so that $E[\bm{I}_c \mid \text{data}]$ is block--diagonal at the MLE, with the $\bm{I}_c^{BB}$ block above and the $\bm{I}_c^{\Sigma\Sigma}$ block as in \eqref{eq:EIc} -- the ordinary Fisher information of an $n$--observation multivariate normal linear model, exactly as if the imputation had never been needed.

\paragraph{The $\text{Var}[\bm{u}_c\mid\text{data}]$ term.} This is where the incompleteness of the data actually enters. Since $\bm{r}_i$ is fixed for $i\in \bm{y}_1$ and, for $i\in \bm{y}_0$, is conditionally Gaussian with mean $\hat{\bm{r}}_i$ and covariance $\bm{V}_i$, the conditional independence of the $\bm{z}_i$ given the data yields three closed--form blocks, built entirely from the $\bm{P}_i$ and $\bm{a}_i$ defined above. Writing $\bm{u}_B$ and $\bm{u}_\Sigma$ for the $\text{vec}(\bm{B})$-- and $\text{vech}(\bm{\Sigma})$--components of $\bm{u}_c$,
\begin{subequations}
\begin{align}
\text{Var}\big[\bm{u}_B \mid \text{data}\big] &= \sum_{i \in \bm{y}_0} \bm{P}_i \otimes \big(\bm{x}_i\bm{x}_i^\top\big), \label{eq:louis-B}\\
\text{Var}\big[\bm{u}_\Sigma \mid \text{data}\big] &= \bm{G}_d^\top\left[ \sum_{i\in \bm{y}_0} \tfrac12\big(\bm{P}_i\otimes \bm{P}_i\big) + \big(\bm{a}_i\bm{a}_i^\top\big)\otimes \bm{P}_i \right] \bm{G}_d, \label{eq:louis-Sigma}\\
\text{Cov}\big[\bm{u}_B,\, \bm{u}_\Sigma \mid \text{data}\big] &= \left[\sum_{i\in \bm{y}_0} \bm{P}_i \otimes \big(\bm{x}_i\bm{a}_i^\top\big)\right] \bm{G}_d. \label{eq:louis-cross}
\end{align}
\end{subequations}
Each of these sums runs only over the incomplete rows $\bm{y}_0$ -- for $i\in\bm{y}_1$, $\bm{r}_i$ is known exactly and contributes no randomness to $\bm{u}_c$ given the data, hence no term to its conditional variance. Unlike $E[\bm{I}_c\mid\text{data}]$, the cross--covariance \eqref{eq:louis-cross} does not vanish at the MLE in general: it is driven by $\bm{a}_i$, i.e.\ by the posterior means $\hat{\bm{r}}_i$ of the imputed residuals, and is zero only in the degenerate case $\hat{\bm{r}}_i = \bm{0}$ for every $i \in \bm{y}_0$.

\paragraph{Assembling $\bm{I}_{obs}$.} Consequently $\bm{I}_{obs}$ is not block--diagonal in $(\bm{B},\bm{\Sigma})$, even though $E[\bm{I}_c\mid\text{data}]$ is: its diagonal blocks follow from \eqref{eq:EIc} minus \eqref{eq:louis-B}--\eqref{eq:louis-Sigma}, and its off--diagonal block is $-\,\text{Cov}[\bm{u}_B,\bm{u}_\Sigma\mid\text{data}]$ from \eqref{eq:louis-cross}. Every term on the right of \eqref{eq:louis} is therefore a function of quantities the E--step has already produced -- $\bm{V}_i$, $\hat{\bm{r}}_i$, together with $\bm{X}$ and $\hat{\bm{\Sigma}}$ -- so that $\bm{I}_{obs}$ can be assembled directly at convergence, with no further optimization. Inverting $\bm{I}_{obs}$ at $(\hat{\bm{B}},\hat{\bm{\Sigma}})$ and reading off the $\text{vec}(\bm{B})$ block then gives asymptotic standard errors for $\hat{\bm{B}}$ at a fraction of the cost of the stratified bootstrap described next; because the cross term \eqref{eq:louis-cross} does not vanish, however, this inversion must use the full joint matrix rather than the $\bm{B}$--block of $\bm{I}_{obs}$ alone, since the latter would ignore the extra uncertainty in $\hat{\bm{B}}$ induced by simultaneously estimating $\hat{\bm{\Sigma}}$ from incomplete data.

Alternatively we may employ stratified non--parametric bootstrap. Since the zero--pattern indicator is treated as ancillary to $(\bm{B},\bm{\Sigma})$, we stratify the bootstrap resampling by observed pattern, resampling with replacement within each stratum at its observed frequency $n_k$. This mirrors the conditioning already implicit in the log--likelihood and avoids the degenerate case of a bootstrap replicate containing zero or one observation of a sparsely--represented pattern, which would otherwise inflate variance for reasons unrelated to genuine parameter uncertainty.

\subsubsection{Residual analysis and effect of zero value patterns}
To assess the adequacy of the CLN and CLNR models we examine the observed part of each compositional vector. For the $i$--th observation with $C_i$ nonzero components, the vector of the alr transformed data $\bm{b}_i=\bm{Q}_i\bm{z}_i$ follows, under the model, a $(C_i-1)$--variate normal distribution with mean $\bm{Q}_i\bm{\mu}_i$ and covariance matrix $\bm{Q}_i\bm{\Sigma}\bm{Q}_i^\top$, where $\bm{\mu}_i=\bm{\mu}$ for the CLN model and $\bm{\mu}_i=\bm{B}^\top\bm{x}_i$ for the CLNR model. Hence the squared Mahalanobis distance
\begin{equation*}
m_i=\left(\bm{b}_i-\bm{Q}_i\bm{\mu}_i\right)^\top
\left(\bm{Q}_i\bm{\Sigma}\bm{Q}_i^\top\right)^{-1}
\left(\bm{b}_i-\bm{Q}_i\bm{\mu}_i\right)
\end{equation*}
follows a $\chi^2_{c_i}$ distribution with $c_i=C_i-1$ degrees of freedom. Since $c_i$ varies with the zero pattern, the distances are not directly comparable across observations. We therefore transform them to normal scores,
$z_i=\Phi^{-1}\left(G_{c_i}(m_i)\right)$, where $G_{c_i}$ is the $\chi^2_{c_i}$ cumulative distribution function, and plot them against the quantiles of the standard normal distribution. Under a correct model the points lie close to the identity line, regardless of the zero pattern. Only the observed alr transformed data are used, so no zero value is replaced or imputed, and the plot uses the same marginal law that enters $\ell_0$ and $\ell_1$ in the likelihood.

The QQ plot therefore checks jointly the assumptions that the alr transformed data are multivariate normal, that the mean structure ($\bm{\mu}$ or $\bm{B}^\top\bm{x}_i$) is correctly specified, and that the covariance matrix $\bm{\Sigma}$ is adequate. Heavy tails or outlying compositions show up as large positive normal scores, whereas a misspecified mean or an underestimated covariance shifts or steepens the whole cloud of points. 

In the resulting plot we colour the observations that contain zeros showing whether they depart from the identity line more than the complete ones, which would suggest that the zero pattern is not independent of $(\bm{\mu},\bm{\Sigma})$, contrary to the assumption. The plot does not assess the model for the zero pattern itself, and because the parameters are estimated from the same data, the $\chi^2$ reference is somewhat optimistic in small samples or when the number of predictors is large.
As a second test of the effect of zero patterns, we may also compare the distributions of the fitted values between the two groups using the non--parametric energy--based test of \cite{sevincc2026}. However, these should be used more as diagnostics and not as formal tests.

\subsection{Miscellanea}
Below we present some straightforward extensions of the CLN and CLNR models.

\begin{itemize}
\item Discriminant analysis may be performed but with caution. If a group contains no zero values, then the standard logistic normal distribution should be fit to this group.
\item The CLNR model can also treat compositional covariates with or without zero values present. If there are no zero values present, the alr transformation may be applied prior to the regression model. In case of zero values present we may rely on the strategy of \cite{alenazi2019} who applied the $\alpha$--transformation \citep{tsagris2011} to the compositional covariates and then used the principal component scores as covariates. 
\item Since the residuals are not of the same dimension, the LASSO variable selection algorithm \citep{tibshirani1996} is not directly applicable and we may rely on alternative algorithms, such as the Forward--Backward with Early Stopping (FBED) \citep{borboudakis2019}.
\item Spatial regression models \citep{tsagris2026} may easily be adopted in this setting. 
\item The CLNR can account for clustered data by using dummy variables to denote the clusters. In case of longitudinal data though, that contain also the time effect, modificatins in the EM algorithm need to be performed.
\end{itemize}

\section{Simulations} \label{sec:sims}
We performed a small scale simulation study to see the effect of the proportion of zero values in the MLE with and without covariates. The simulations were carried over using a Dell laptop with i7--1355U processor, 16GB of RAM and windows installed, and the \textit{R} code to reproduce the simulations is available in the supplementary material. Prior to these, we produce some contour plots of the CLN and ZAD models. 

\subsection{Visualization of the CLN}
We visualize the densities of the CLN and the ZAD models to show their differences. At first, we generated 150 observations from a Dirichlet distribution with parameters $a=c(2,4,8)$ and selected 10 observations at random and zeroed 1 component at each. We then normalized those observations to map on the simplex and estimated the CLN and ZAD models. We then generated 150 observations from a bivariate normal distribution and mapped the values onto the simplex using the inverse of the alr transformation (\ref{alrinv}). We zeroed again 10 components using the aforementioned procedure. 

\begin{figure}[!ht]
\centering
\begin{tabular}{cc}
\multicolumn{2}{c}{\textbf{Dirichlet is the ground truth model}.} \\
\includegraphics[scale = 0.45, trim = 80 0 0 0]{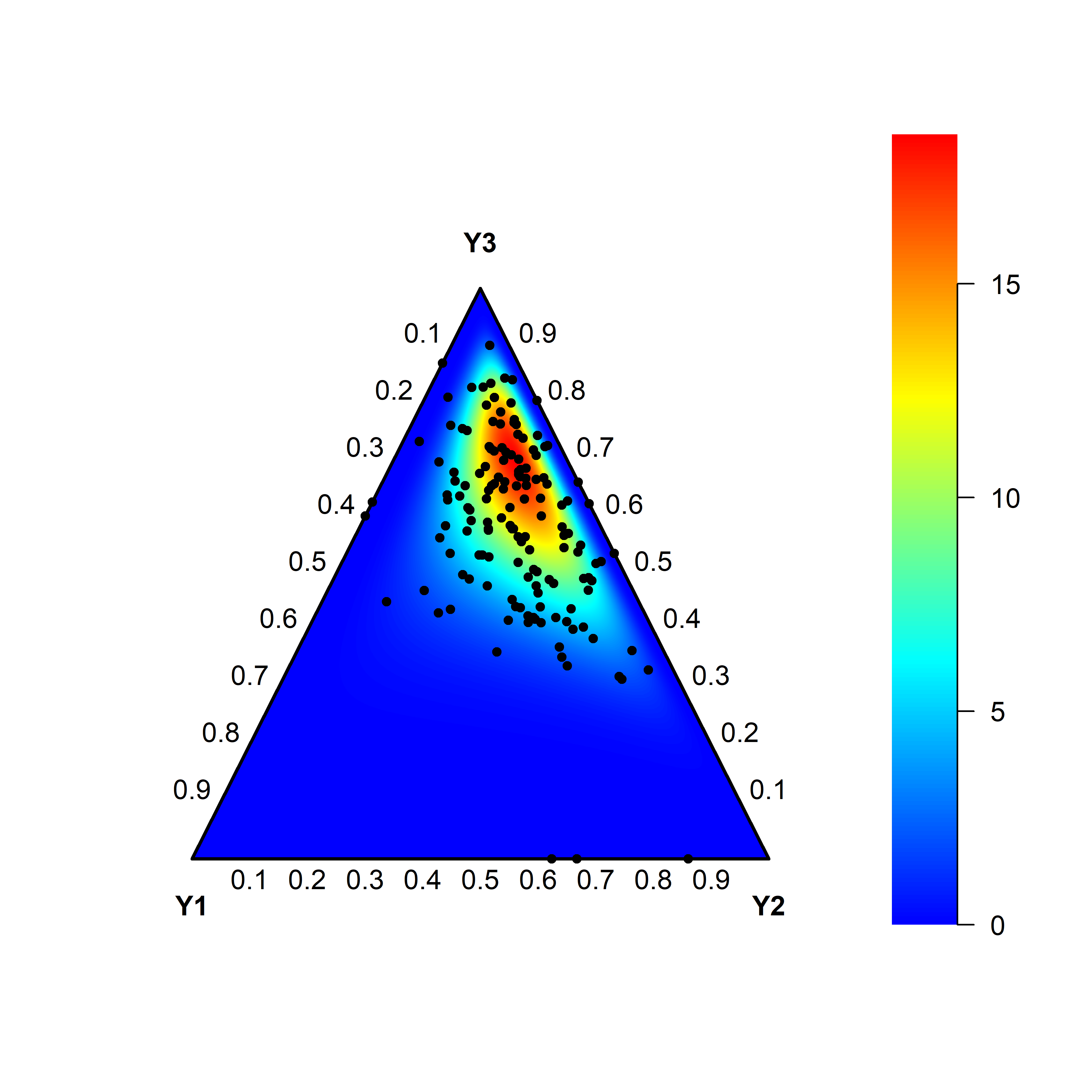} &
\includegraphics[scale = 0.45, trim = 60 0 0 0]{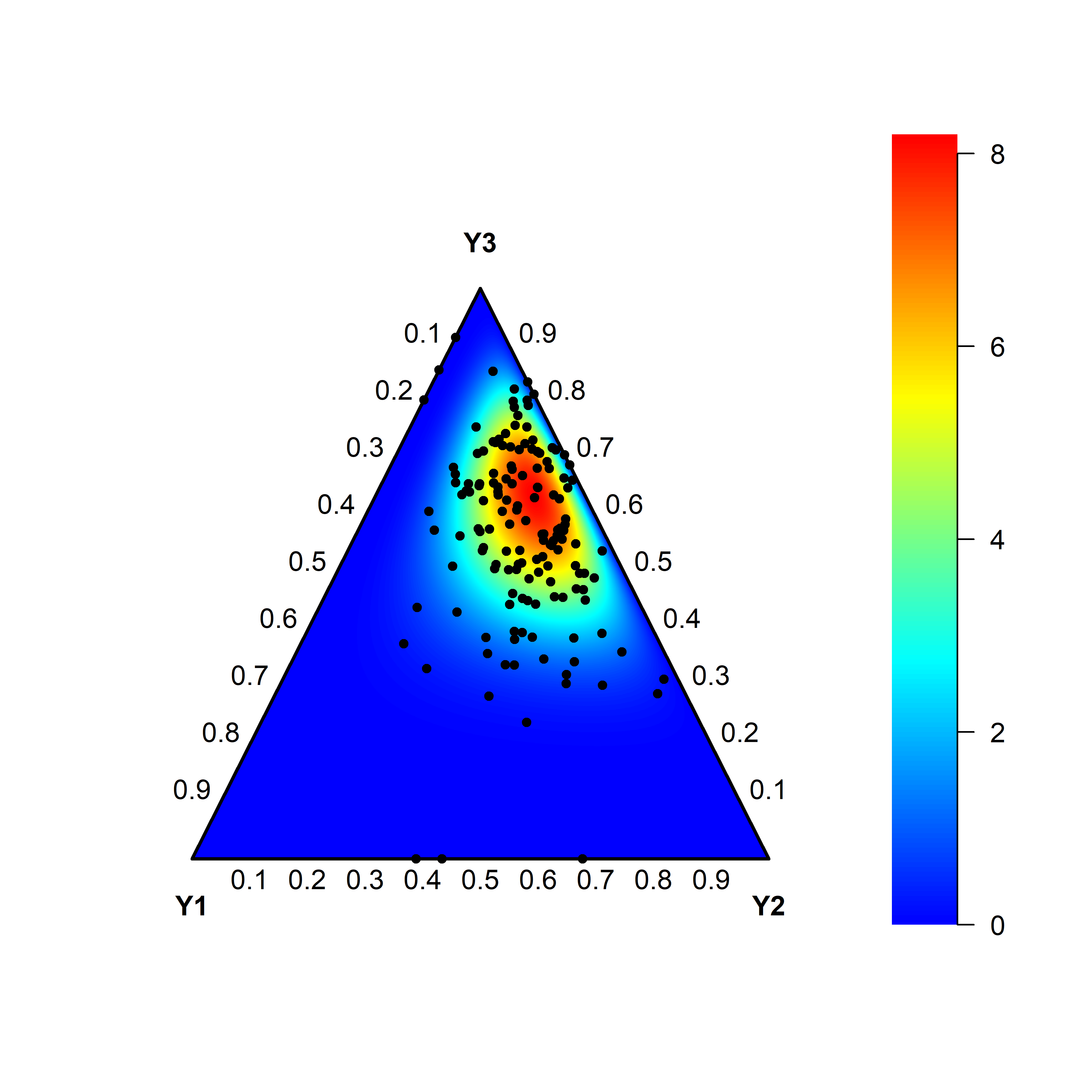} \\
(a) Contour plot of the estimated ZAD model. & (b) Contour plot of the estimated CLN model. \\
\multicolumn{2}{c}{\textbf{Normal is the ground truth model}.} \\
\includegraphics[scale = 0.45, trim = 80 0 0 0]{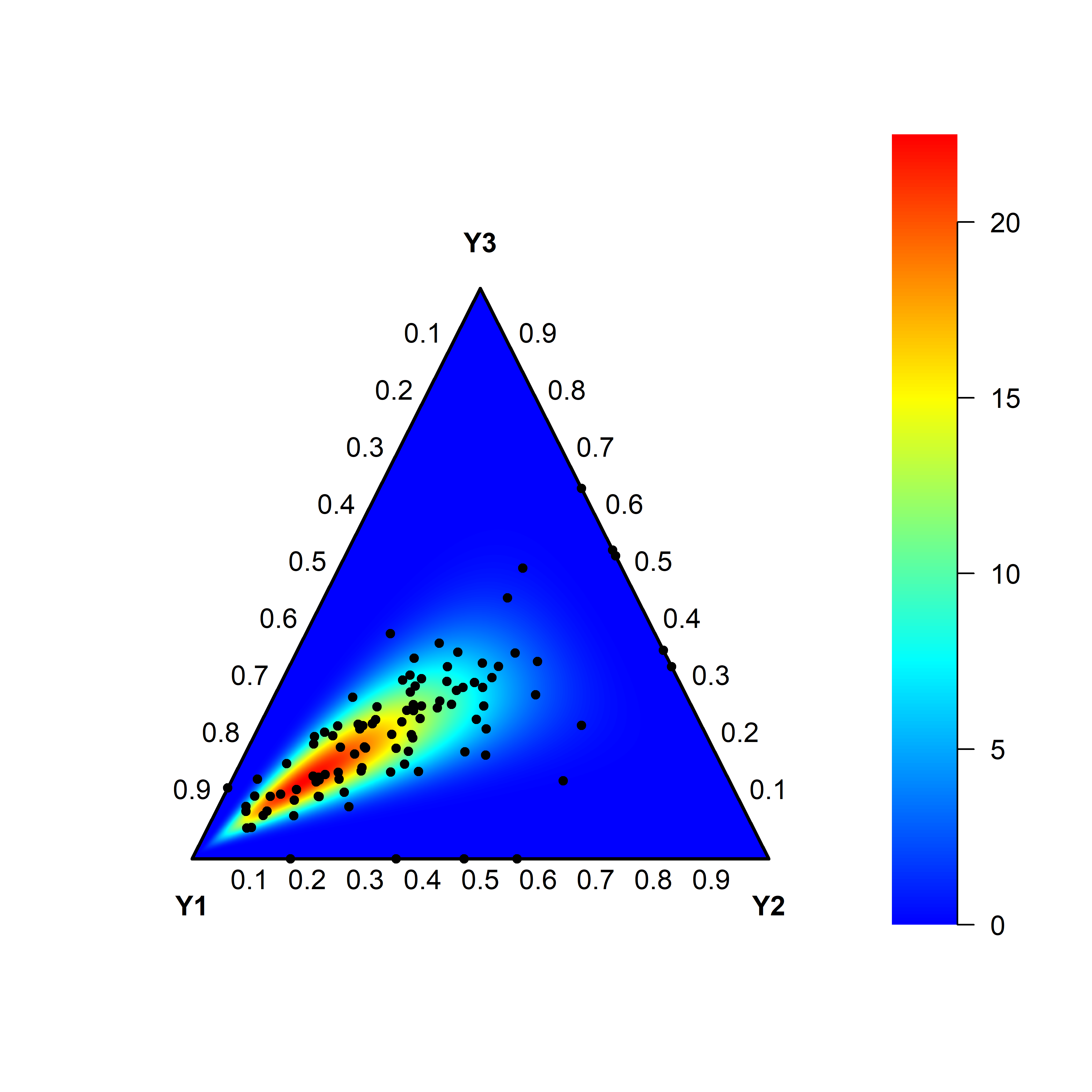} &
\includegraphics[scale = 0.45, trim = 60 0 0 0]{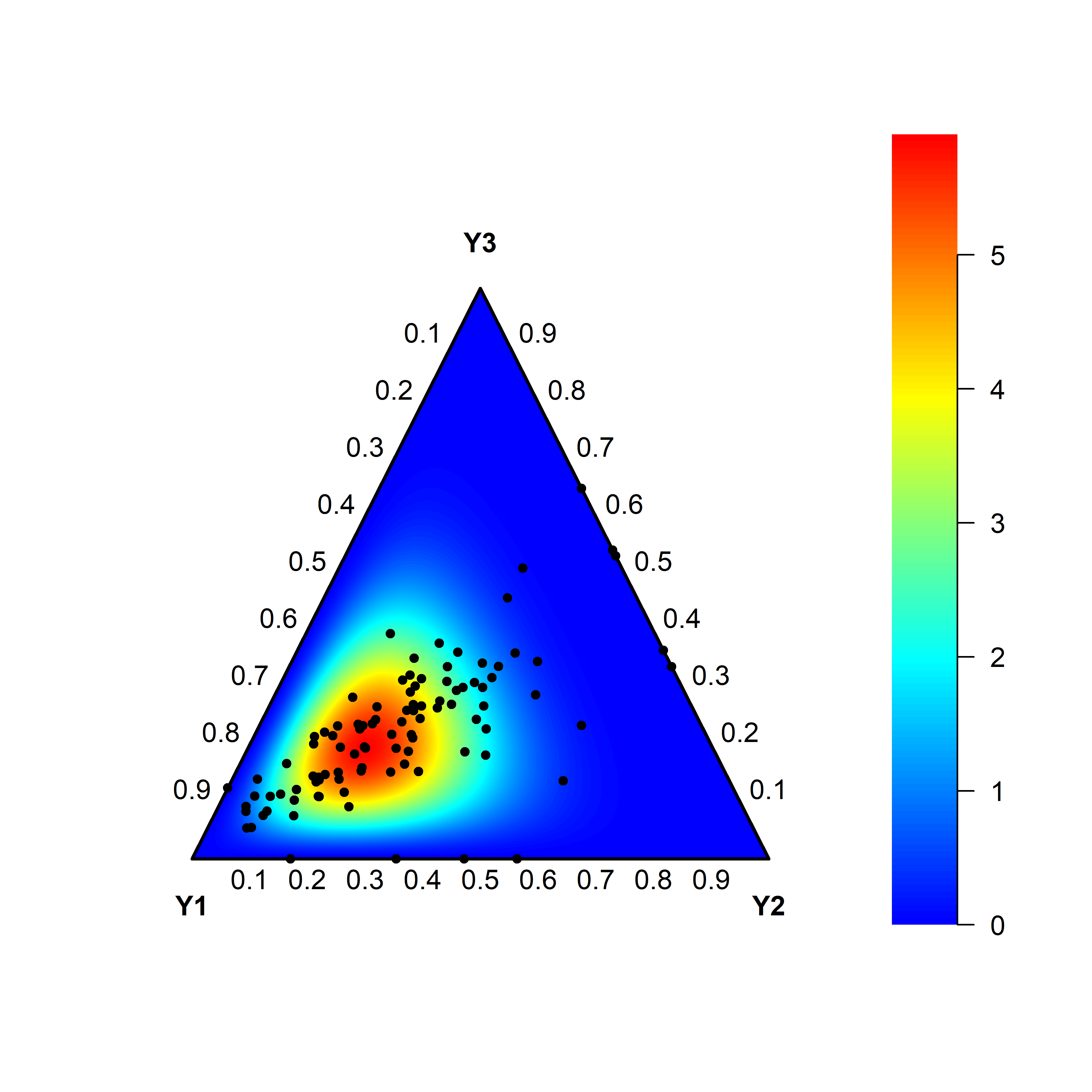} \\
(c) Contour plot of the estimated ZAD model. & (d) Contour plot of the estimated CLN model.
\end{tabular}
\caption{Estimated discrepancies between the fitted and the true normal distribution. On the left are the distances between the mean vectors, and on the right, the distances between the covariance matrices.} 
\label{contours}
\end{figure}

Figure \ref{contours} presents the contour plots of the estimated CLN and ZAD models. The first row contains the contours when the Dirichlet distribution is the true model, and the second row the contours when the normal is the true model. Evidently, CLN captures the Dirichlet shape, while the inverse is not true. This result is not new and was noted by \cite{ait2003}, and it shows the effectiveness of the CLN compared to ZAD. 

\subsection{Performance of the CLN model}
We generated data of sample sizes $n=(100,200,\ldots,1000)$ from a $N_4\left(\bm{\mu},\bm{\Sigma}\right)$, where $\bm{\mu}$ and $\bm{\Sigma}$ are the mean vector and covariance matrix, respectively, and then mapped the data back onto the simple using the inverse of the alr transformation (\ref{alrinv}). We then selected at random $q=(10\%,\ldots,50\%)$ vectors for each sample size and zeroed, at random, 2 values, and renormalized the data to lie within the simplex. We fitted the CLN model in the simulated data and compute the Kullback--Leibler divergence (KLD) between the fitted normal and the true normal distribution. We then aggregated the divergences over 500 repetitions.  

\begin{figure}[!ht]
\centering
\includegraphics[scale = 0.45, trim = 80 0 0 0]{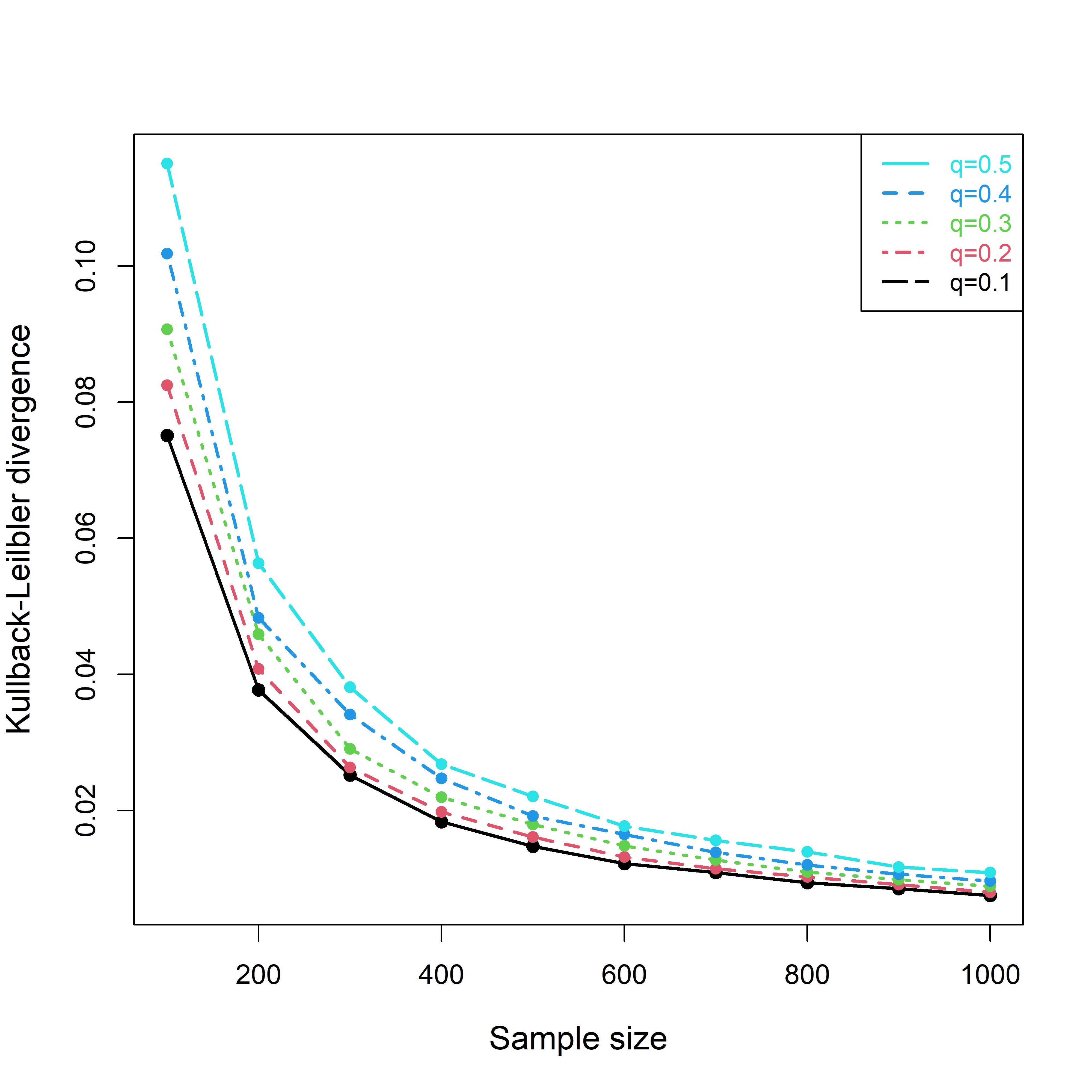}
\caption{KLD between the estimated CLN distribution and the true normal distribution.} 
\label{kl}
\end{figure}

Figure \ref{kl} presents the results. Evidently, as the proportion of compositional vectors with zeros increases, so does the KLD. The decreasing pattern though is the same for all proportions of zeros. 

\subsection{Type I error of the James test}
We generated data for two groups, $\bm{y}_1$ and $\bm{y}_2$, both from a logistic normal distribution with $d=4$ components. The sample sizes considered were $(50, 100, 150, 200)$, and  we used the following parameters
\begin{eqnarray*}
\bm{\mu}=\left(5.843, 3.057, 3.758, 1.199\right)^\top \ \ \text{and} \ \
\bm{\Sigma}=\left[
\begin{array}{rrrr}
 0.686 & -0.042 &  1.274  & 0.516  \\
-0.042 &  0.190 &  -0.330 & -0.122 \\
1.274  & -0.330 &  3.116  & 1.296  \\
0.516  & -0.122 &  1.296  & 0.581 
\end{array} \right].
\end{eqnarray*}
The first sample ${y}_1$ we generated $n_1$ observations from a multivariate logistic normal $MLN(\bm{\mu},\bm{\Sigma})$ whereas the second sample $\bm{y_2}$ was generated from a multivariate logistic normal $MLN(\bm{\mu}, 5\bm{\Sigma})$. For each combination of sample sizes we generated data and performed the James test, using 1,000 replicates, and computed the proportion of times the null hypothesis of equal mean vectors was rejected at the $\alpha=0.05$ significance level. 

Table \ref{jamesalpha} presents the estimated type I error under these settings. Evidently, even if one group has moderate sample size, the test is not size correct. We can observe that both ample sizes must be larger than $\approx 150$ for the test to attain the correct type I error. Bootstrap calibration is not straightforward under the CLN model, and for this we also implemented the permutation--based version of the James test, but the results remained unaltered.

\begin{table}[ht]
\centering
\caption{Estimated type I error of the James test with unequal covariance matrices. $n_1$ and $n_2$ denote the sample sizes of the first and the second group, respectively.}
\label{jamesalpha}
\begin{tabular}{lrrrr}
\toprule
          & $n_2=50$ & $n_2=100$ & $n_2=150$ & $n_2=200$ \\ \midrule
$n_1=50$  & 0.068 & 0.068 & 0.067 & 0.075 \\ 
$n_1=100$ & 0.058 & 0.058 & 0.056 & 0.062 \\ 
$n_1=150$ & 0.061 & 0.063 & 0.050 & 0.058 \\ 
$n_1=200$ & 0.066 & 0.073 & 0.049 & 0.053 \\ 
\bottomrule
\end{tabular}
\end{table}

\subsection{Performance of the CLNR model}
We used  a design matrix $\bf X$, that composes of 1 or 3 predictor variables,  generated from a normal distribution with the response compositional data $(\bf Y)$ consisting of $D = (4, 7, 10)$ components. At first, Euclidean data were generated as 
\begin{eqnarray} \label{mu}
\mu_i=\left({\bf x}_1^T\pmb{\beta}_1+e_1,\ldots, {\bf x}^T\pmb{\beta}_d+e_d \right)^\top,
\end{eqnarray}
where $i=1,\ldots,n$, $e_j \sim N(0, 1), \ \ j = 2, \ldots, d$. The matrix of regression coefficients ${\bf B}=\left(\pmb{\beta}_1^T, \ldots, \pmb{\beta}_d^T\right)$ was generated from a normal distribution, with the constant terms being generated from $N(-2, 1)$ and the slopes from $N(2, 1)$. The data were then mapped onto the simplex using the inverse of the alr transformation (\ref{alrinv}). We selected at random $25\%$ of the observations and for each compositional vector, we set $40\%$ of their components equal to zero. For instance, when $D=4$, 1 component was set to zero, and when $D=7$ there were 2 components whose values were zeroed, and so on. Those vectors were again normalized to sum to 1. The data were intentionally not generated from the true CLN model since the CLN is tested and the goal was to examine its behavior under some degree of model miss--specification. This time, we considered sample sizes of $n=(100, 300, 500, 1000, 2000, 3000, 5000)$.

The bias of the estimated coefficients of either regression model was computed via the Frobenius norm $\left\|\widetilde{\bf B}-{\bf B}\right\|_F$ averaged over 500 repetitions, where ${\bf B}$ and $\widetilde{\bf B}$ denotes the true and the estimated regression coefficients, respectively. We also measured the goodness of fit via the KLD between the observed and the fitted compositions. 

Table \ref{bias1} presents the results of the simulations. Evidently, the estimated bias of the regression coefficients produced the CLNR model is significantly lower than that of the ZAD regression (ZADR) model. Regarding the goodness of fit, the CLN outperforms ZADR, especially in the large sample sizes. 

\begin{table}[ht]
\centering
\begin{footnotesize}
\caption{Estimated bias of the regression coefficients and KLD between the observed and fitted compositions, with 1 and 3 predictor variables.}
\label{bias1}
\begin{tabular}{l|cccccc|cccccc}
\toprule
       & \multicolumn{6}{c}{One predictor} & \multicolumn{6}{c}{3 Predictors} \\ \midrule
       &  \multicolumn{12}{c}{Estimated bias} \\ \midrule
Sample & \multicolumn{2}{c}{D=4} & \multicolumn{2}{c}{D=7} & \multicolumn{2}{c}{D=10} 
       & \multicolumn{2}{c}{D=4} & \multicolumn{2}{c}{D=7} & \multicolumn{2}{c}{D=10} \\ \midrule
size   & CLNR & ZADR & CLNR & ZADR & CLNR & ZADR & CLNR & ZADR & CLNR & ZADR & CLNR & ZADR \\  \midrule
$n=100$  & 0.215 & 0.207 & 0.308 & 3.244 & 0.361 & 2.556 & 0.292 & 7.664 & 0.437 & 15.007 & 0.527 & 20.588 \\ 
$n=300$  & 0.121 & 0.143 & 0.181 & 3.263 & 0.203 & 2.613 & 0.166 & 7.836 & 0.239 & 15.397 & 0.319 & 20.938 \\ 
$n=500$  & 0.090 & 0.129 & 0.131 & 3.286 & 0.166 & 2.643 & 0.132 & 7.830 & 0.191 & 15.425 & 0.233 & 21.056 \\ 
$n=1000$ & 0.064 & 0.126 & 0.098 & 3.293 & 0.114 & 2.610 & 0.095 & 7.861 & 0.139 & 15.500 & 0.174 & 21.098 \\ 
$n=2000$ & 0.048 & 0.122 & 0.066 & 3.298 & 0.082 & 2.635 & 0.067 & 7.875 & 0.094 & 15.497 & 0.111 & 21.100 \\  \midrule 
        & \multicolumn{12}{c}{KLD} \\ \midrule
        & \multicolumn{2}{c}{D=4} & \multicolumn{2}{c}{D=7} & \multicolumn{2}{c}{D=10}
        & \multicolumn{2}{c}{D=4} & \multicolumn{2}{c}{D=7} & \multicolumn{2}{c}{D=10} \\ \midrule
$n=100$ & 24.773 & 22.181 & 31.482 & 29.690 & 35.034 & 35.871 & 24.364 & 23.007 & 31.396 & 30.838 & 33.897 & 36.320 \\
$n=300$ & 75.960 & 67.961 & 96.187 & 90.735 & 106.319 & 109.083 & 74.142 & 70.857 & 93.508 & 94.796 & 103.638 & 113.010 \\ 
$n=500$ & 128.295 & 114.387 & 160.843 & 152.076 & 177.909 & 182.625 & 124.098 & 118.700 & 156.989 & 159.566 & 173.479 & 189.359 \\
$n=1000$ & 254.528 & 227.834 & 323.102 & 304.939 & 356.286 & 366.169 & 247.893 & 237.820 & 313.246 & 319.687 & 347.213 & 380.646 \\
$n=2000$ & 511.776 & 457.497 & 647.624 & 611.500 & 711.851 & 731.984 & 498.231 & 477.579 & 627.887 & 641.580 & 694.687 & 762.407  \\ 
\bottomrule
\end{tabular}
\end{footnotesize}
\end{table}

\subsection{Type I error of the CLNR model} 
We simulated compositional responses from a logistic normal distribution of dimensions $D=4, 7, 10$ and values for 3 predictor variables from the standard normal distribution. We then selected 25\% observations and zeroed at random 1, 2, and and 3 components when $D=4, 7$ and $10$, respectively. The sample sizes considered were $n=100, 300, 500, 1000$. For each combination of $n$ and $D$ the log--likelihood ratio test using the CLNR and ZADR models to assess the significance of the regression coefficients, was performed 1,000 times, and the proportion of rejections at the $5\%$ significance level was computed. 

Table \ref{typeI} shows that with the exception of $D=10$ and $n=100$, the CLNR model is size correct. ZADR on the other hand, when $D=10$, is anti--consrvative, regardless of the sample size. 

\begin{table}[ht]
\centering
\caption{Estimated type I error for the CLNR and the ZADR models.}
\label{typeI}
\begin{tabular}{l|rrr|rrr}
\toprule
         & \multicolumn{3}{c}{CLNR} & \multicolumn{3}{c}{ZADR} \\ \midrule
         & D=4 & D=7 & D=10 & D=4 & D=7 & D=10 \\  \midrule
$n=100$  & 0.066 & 0.073 & 0.103 & 0.042 & 0.062 & 0.100 \\ 
$n=300$  & 0.057 & 0.067 & 0.061 & 0.027 & 0.050 & 0.106 \\ 
$n=500$  & 0.052 & 0.056 & 0.049 & 0.029 & 0.049 & 0.099 \\ 
$n=1000$ & 0.049 & 0.058 & 0.046 & 0.033 & 0.052 & 0.127 \\ 
\bottomrule
\end{tabular}
\end{table}

\subsection{Computational cost}
We then compared the computational cost of the CLN to ZAD with and without covariates. Following the same data generation process but using the same sample size as in the simple MLE case we measured the running time to fit each model\footnote{These timings were measured in a Dell laptop with Intel Core i7--1355U processor, 16GB RAM, 500 GB SSD and Windows OS, using \textit{R} 4.6.1. Both functions are fully implemented in \textit{R}.}, repeating this process 20 times and averaging the times over these 20 repetitions. For the ZAD and ZADR models we used iteratively reweighted least squares (IRLS) algorithm that is significantly faster than \textit{R}'s built--in \textit{optim()} function used by \cite{tsagris2018}.

\begin{figure}[!ht]
\centering
\includegraphics[scale = 0.45, trim = 80 0 0 0]{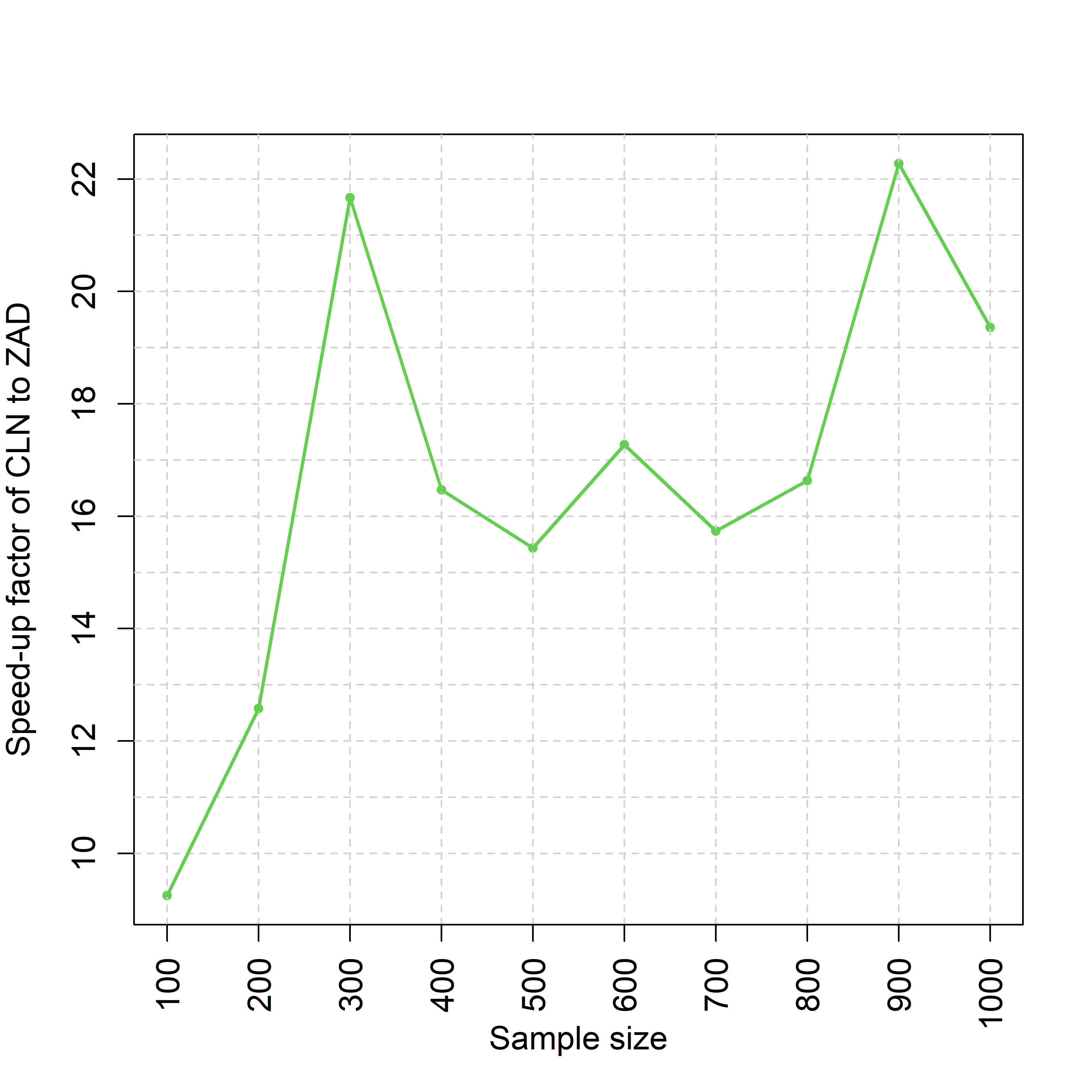}
\caption{Speed--up factor of the CLN MLE to ZAD MLE.} 
\label{mle.time}
\end{figure}

Figure \ref{mle.time} visualizes the speed--up factors\footnote{Ratio of the computanational cost (running time) of ZAD to CLN.} of the CLN MLE to ZAD MLE versus the sample size. The speed--up factors range from 9.3 up to 22.3, with an average equal to 16.7. Table \ref{times} showcases the computational cost (in seconds) of each regression model. On average, with 1 predictor, the speed--up factors of the CLNR model compared to ZADR are 12.5 when $D=4$, 18.6 when $D=7$ and 18.9 when $D=10$. The average speed--up factor with 3 predictors are 22.5 when $D=4$, 25.6 when $D=7$ and 30 when $D=10$.

\begin{table}[ht]
\centering
\begin{small}
\caption{Computational cost (in seconds) of the CLNR to ZADR model with 1 and 3 predictor variables.}
\label{times}
\begin{tabular}{r|rrrrrr|rrrrrr}
\toprule
 & \multicolumn{6}{c|}{1 predictor variable} & \multicolumn{6}{c}{3 predictor variables} \\ \midrule
Sample & \multicolumn{2}{c}{$D=4$} & \multicolumn{2}{c}{$D=7$} & \multicolumn{2}{c|}{$D=10$} & \multicolumn{2}{c}{$D=4$} & \multicolumn{2}{c}{$D=7$} & \multicolumn{2}{c}{$D=10$} \\ \midrule
size & CLNR & ZADR & CLNR & ZADR & CLNR & ZADR & CLNR & ZADR & CLNR & ZADR & CLNR & ZADR \\ \midrule
$n=100$  & 0.01 & 0.07 & 0.02 & 0.31 & 0.04 & 0.45 & 0.01 & 0.14 & 0.03 & 0.53 & 0.04 & 0.74 \\
$n=200$  & 0.04 & 0.40 & 0.05 & 0.84 & 0.06 & 0.95 & 0.04 & 0.87 & 0.06 & 1.14 & 0.06 & 1.44 \\
$n=300$  & 0.06 & 0.67 & 0.07 & 1.19 & 0.09 & 1.37 & 0.06 & 1.17 & 0.07 & 1.68 & 0.07 & 2.09 \\
$n=400$  & 0.09 & 1.00 & 0.08 & 1.58 & 0.10 & 1.91 & 0.08 & 1.70 & 0.09 & 2.27 & 0.08 & 2.61 \\
$n=500$  & 0.10 & 1.16 & 0.12 & 2.19 & 0.12 & 2.37 & 0.09 & 2.02 & 0.10 & 2.64 & 0.12 & 3.75 \\
$n=600$  & 0.09 & 1.24 & 0.12 & 2.44 & 0.14 & 2.98 & 0.11 & 2.47 & 0.13 & 3.55 & 0.15 & 4.84 \\
$n=700$  & 0.13 & 1.67 & 0.13 & 2.53 & 0.15 & 3.10 & 0.13 & 3.01 & 0.15 & 4.18 & 0.17 & 5.42 \\
$n=800$  & 0.13 & 1.78 & 0.16 & 3.12 & 0.19 & 3.62 & 0.14 & 3.36 & 0.16 & 4.68 & 0.20 & 6.32 \\
$n=900$  & 0.15 & 1.87 & 0.15 & 3.21 & 0.17 & 3.99 & 0.17 & 3.93 & 0.18 & 5.13 & 0.20 & 6.79 \\
$n=1000$ & 0.16 & 2.12 & 0.17 & 3.83 & 0.21 & 4.55 & 0.08 & 2.17 & 0.19 & 5.84 & 0.23 & 7.63 \\ \bottomrule
\end{tabular}
\end{small}
\end{table}

\section{Real data analysis} \label{sec:real}
The first dataset will be used to illustrate the performance of the CLN based PCA as a dimensionality reduction method, whereas we will illustrate the performance of the CLNR model and compare it to ZADR using two real datasets. For one of them we will also dispaly the CLN based biplot of the data. 

\subsection{Honey dataset}
\begin{figure}[!ht]
\centering
\includegraphics[scale = 0.55]{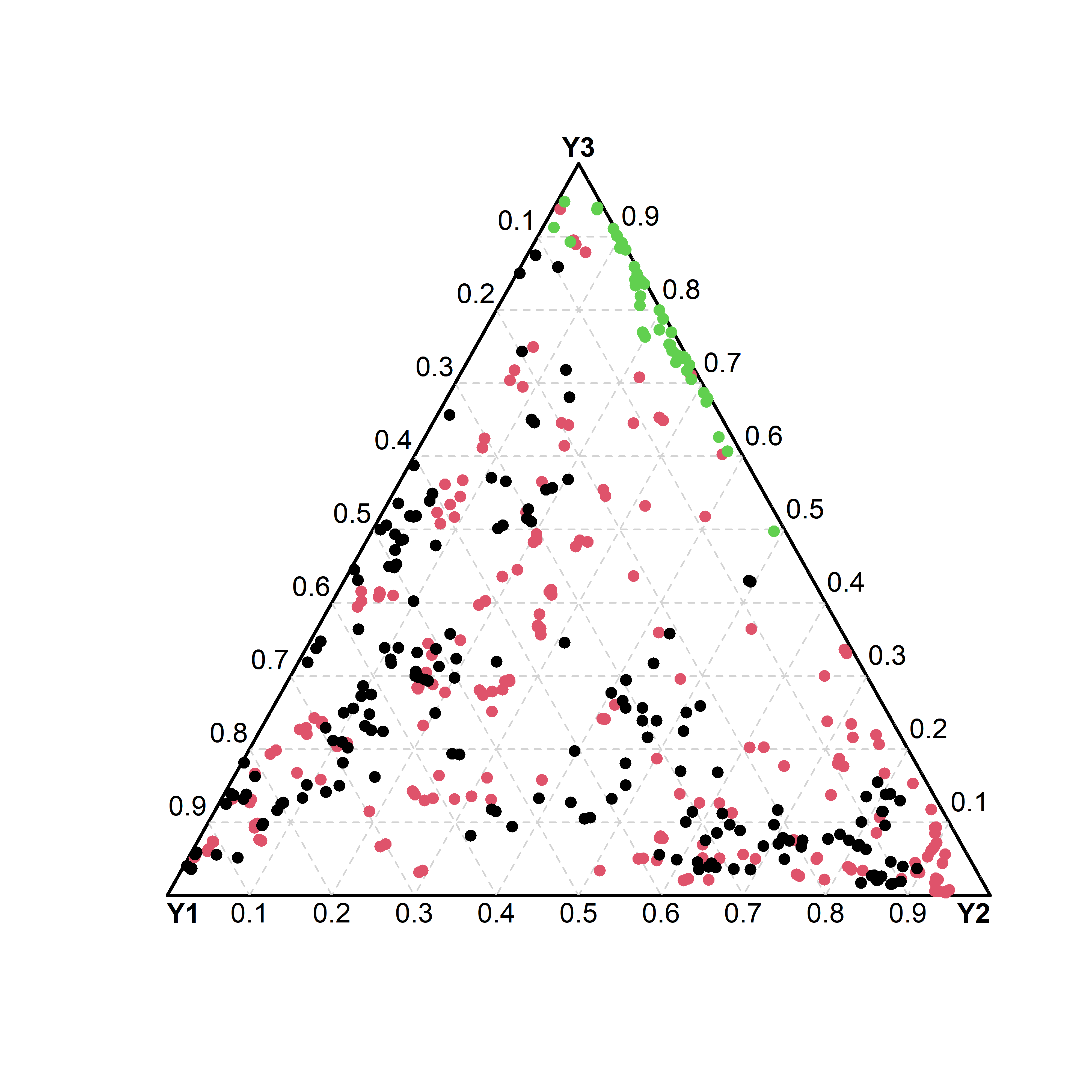} 
\caption{Ternary plot of the PCA applied to the Honey dataset reducing its dimensionality to 3 components.} 
\label{honey}
\end{figure}

The honey dataset contains 429 compositional vectors and 12 components. It contains information on three types of honey, adulterated honey, honey or syrup, and the measurements in 12 chemical elements. We will use this dataset as an illustration of the PCA, reducing the dimensinality two 2 variables, or a 3--component composition. Figure \ref{honey} is the ternary plot visualizing the dimensionality reduced compositional data, where the colour indicates the different types of honey. We can see that two types of honey are not easily distinguished, whereas one type of honey is well separated, explaining why the simplicial clustering of \cite{tsagris2025b} failed to identify the three types.

\subsection{Glacial dataset}
The glacial dataset contains the percentages by weight in 92 observations of pebbles of glacial tills of four categories (red sandstone, gray sandstone, crystalline and miscellaneous). The interest lies in relating the compositions to the total pebbles counts. The data set is available in the \textit{R} package \textit{compositions} \citep{compositions2025} and 42 observations at least one zero value and the logarithm of the counts was used as a covariate.

Table \ref{reg1} contains the estimated regression coefficients of the CLN and the ZADR models. The red sandstone (first component) was chosen as the common divisor in the alr transformation. Evidently, the slopes for the $\log{\left(\text{Crytalline/Red sandstone}\right)}$ and the $\log{\left(\text{Miscellanea/Red sandstone}\right)}$ are opposite between the two models. Note also that the slope of the crystalline component is large compared to its associated error, for the CLNR model. A log--likelihood ratio test reveals that this covariate is significant based on the CLNR (p--value=0.001), whereas based on ZADR, the covariate is not significant (p--value=0.086).

\begin{figure}[!ht]
\centering
\begin{tabular}{cc}
\includegraphics[scale = 0.45, trim = 60 0 0 0]{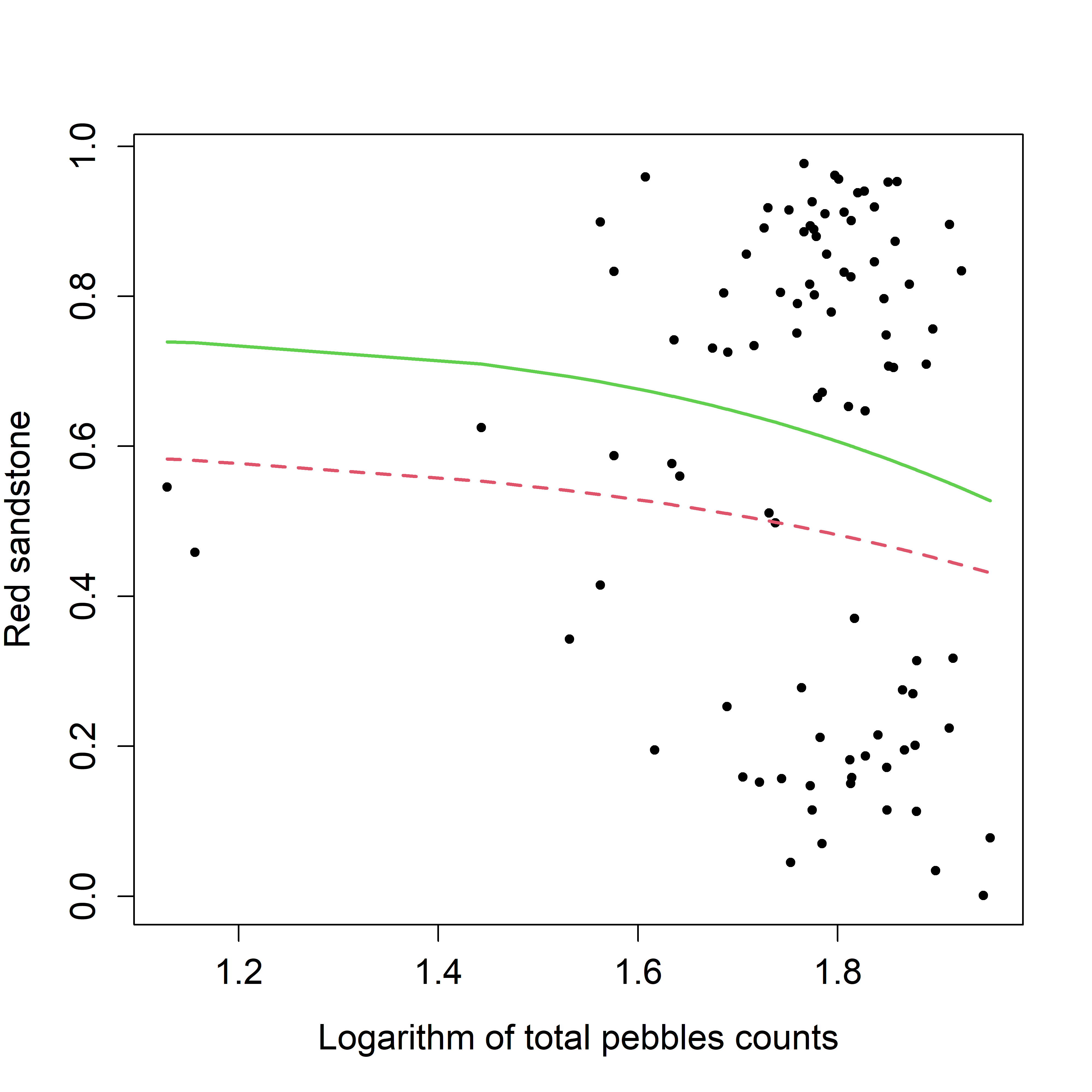} &
\includegraphics[scale = 0.45, trim = 0 0 0 0]{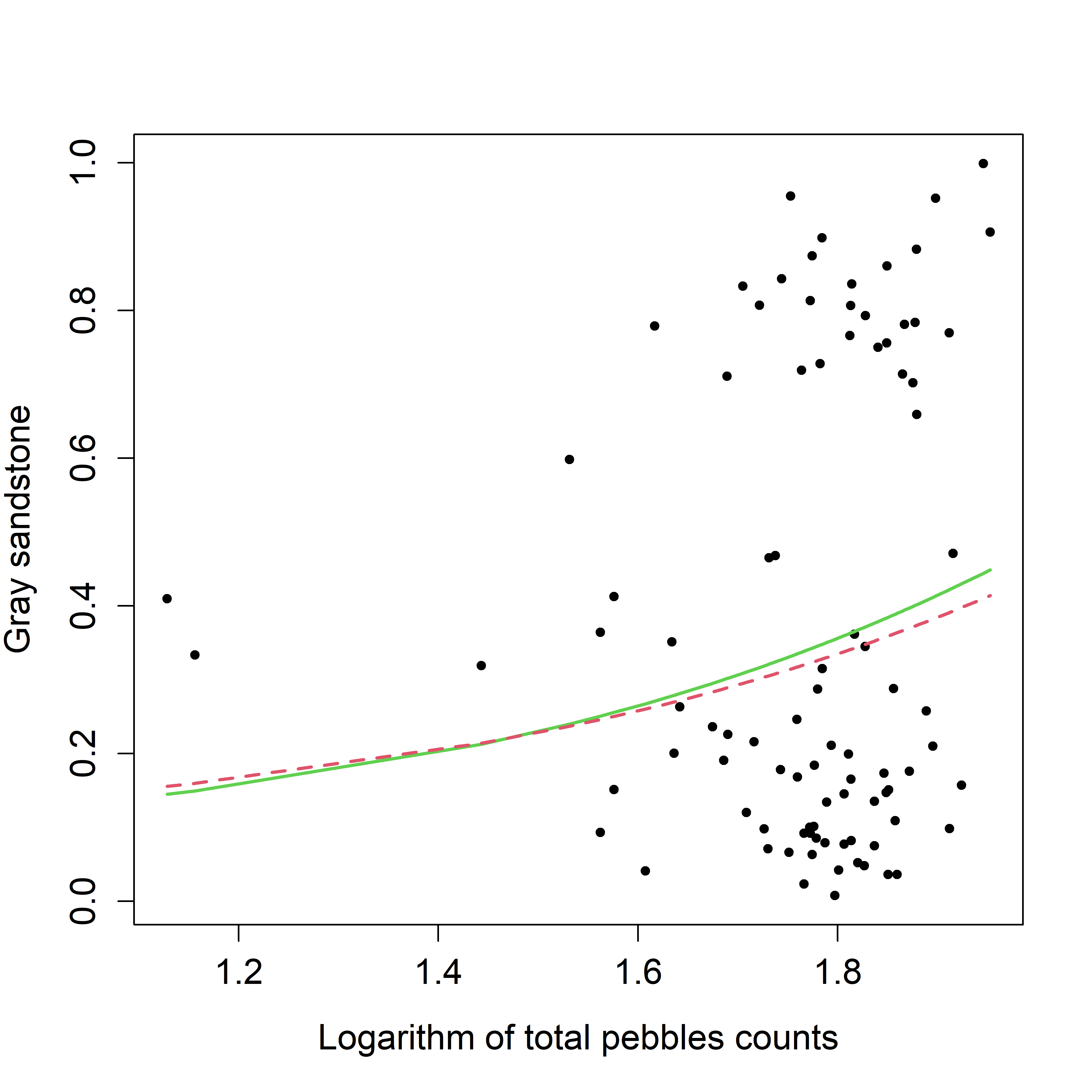} \\
(a) Red sandstone. & (b) Gray sandstone. \\
\includegraphics[scale = 0.45, trim = 60 0 0 0]{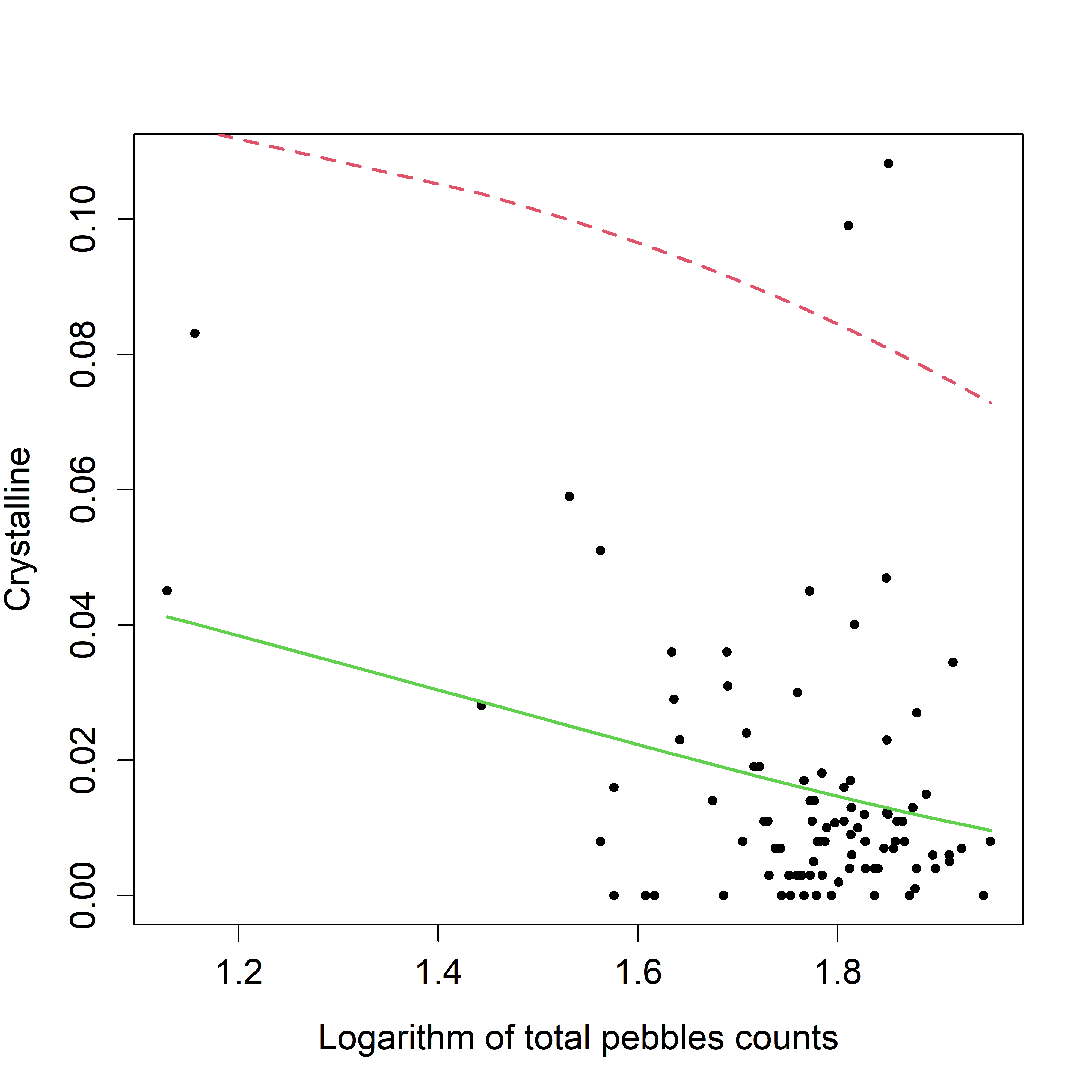} &
\includegraphics[scale = 0.45, trim = 0 0 0 0]{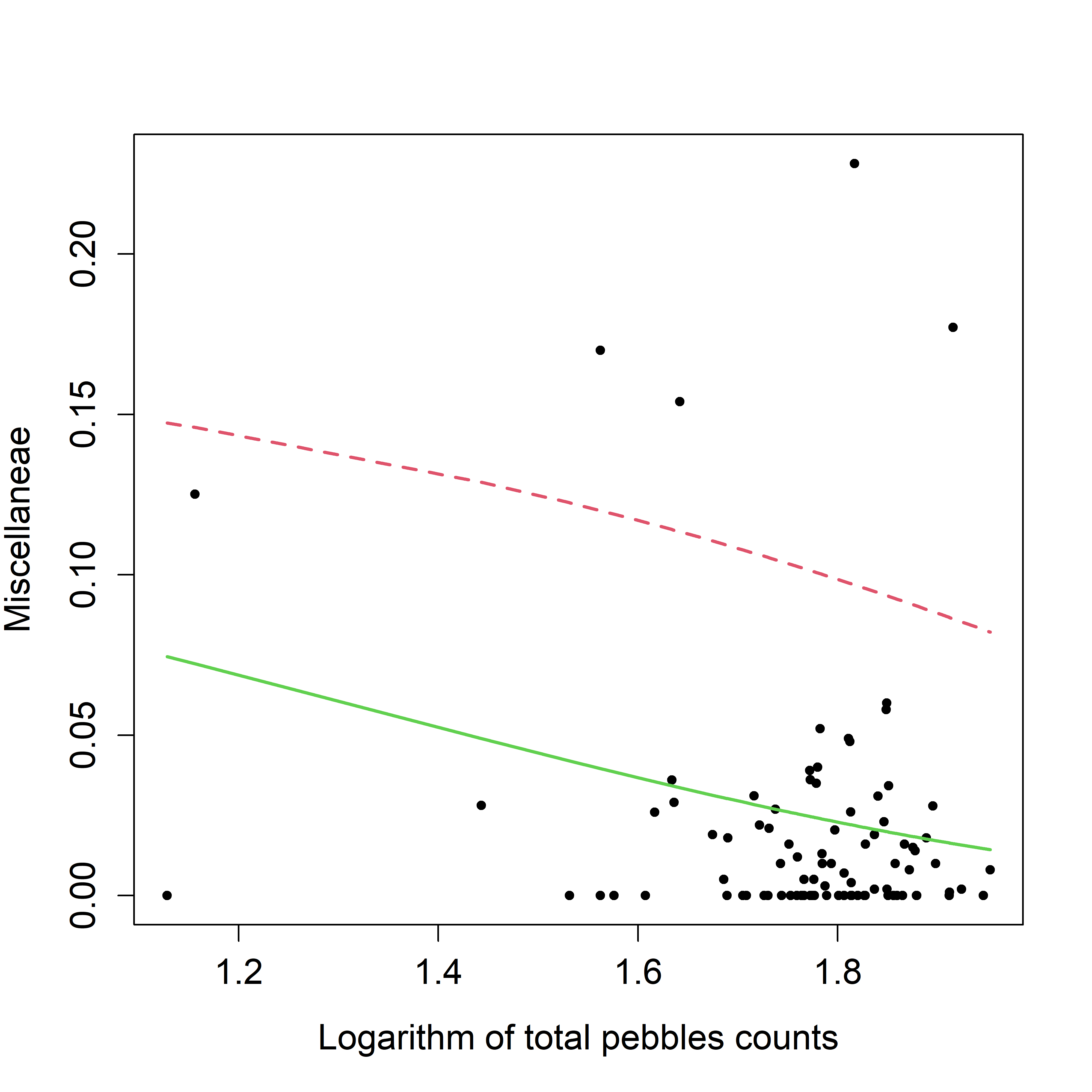} \\
(c) Crystalline. & (d) Miscellanea.
\end{tabular}
\caption{Logarithm of the total pebble counts versus the observed and the fitted compositions. The green solid line indicates the fitted values of the CLNR model, whereas the red dotted line indicates the fitted values of the ZADR model.} 
\label{glass}
\end{figure}

Figure \ref{glass} contains the scatter plots of the logarithm of the total pebble counts versus the observed and fitted compositions for each component. Only for the Gray sandstone the fitted values are close between the two models. For the crystalline and the Miscellanea components it is evident that the ZADR model is heavily affected by some values far from the bulk of the data. 
Finally, the KLD between the observed and the fitted compositions are 23.92 and 33.29 for the CLNR and ZADR models, respectively, indicating that the CLN has a better fit. 

\begin{figure}[!ht]
\centering
\includegraphics[scale = 0.45, trim = 0 0 0 0]{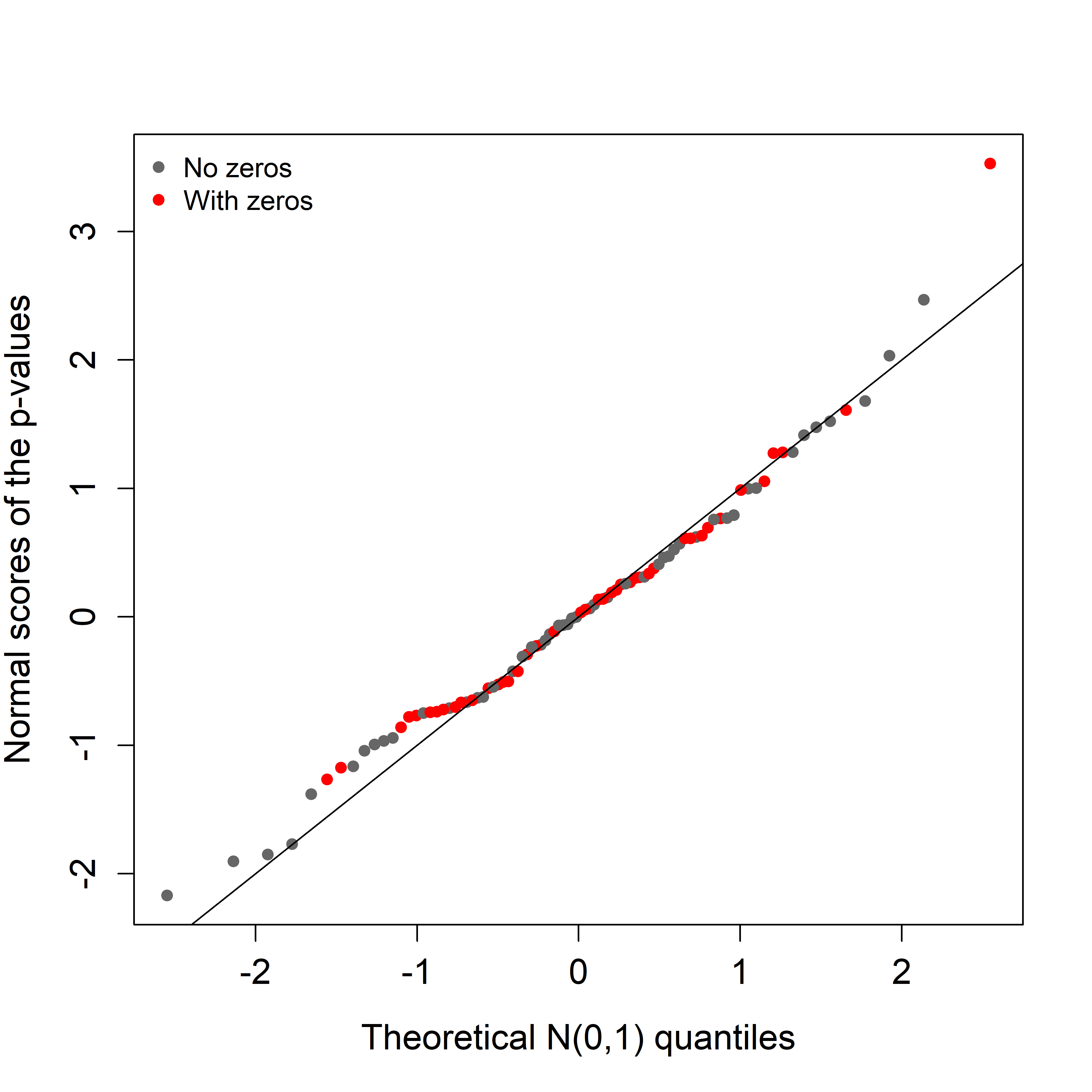} \\
\caption{QQ plot of the residuals of the CLNR model applied to the Glacial dataset.} 
\label{resid1}
\end{figure}

Figure \ref{resid1} contains the QQ plot of the CLNR model. The red and grey points, corresponding to compositions with and without zero values present, respectively seem uniformly scattered within the bulk of the points. Further, the non--parametric energy--based test of distributions provided a $p-value$ equal to 0.998.

\begin{table}[ht]
\centering
\caption{Regression coefficients of the CLNR and ZADR models applied to the Glacial dataset. Their estimated standard errors appear within parentheses. For the CLNR we provide, for comparison purposes, first, the theoretical estimate (\ref{eq:louis}), and second, the bootstrap based estimate. For the ZADR, the standard error is produced by the observed information matrix.}
\label{reg1}
\begin{tabular}{l|rrr}
\toprule
& Gray sandstone & Crystalline & Miscellanea \\   \midrule
          & \multicolumn{3}{c}{CLNR} \\ \midrule
Intercept & 1.269(1.626, 1.694)  & -1.507(1.359, 1.379) & -0.747(1.323, 1.185) \\ 
Slope     & 0.332(0.273, 0.294)  & 0.703(0.228, 0.239)  & 0.050(0.221, 0.206) \\ \midrule
          & \multicolumn{3}{c}{ZADR} \\ \midrule
Intercept & -2.321(1.014) & -1.518(1.057) & -1.154(1.304) \\ 
Slope     & 0.323(0.169)  & -0.037(0.178) & -0.071(0.219) \\  \bottomrule
\end{tabular}
\end{table}

\subsection{Elections dataset}
The elections dataset contains information on the 2000 U.S. presidential election in the 67 counties of Florida. For each county, the percentage of votes of each of the 10 candidates is available along with information on 8 predictor variables, such as population, percentage of population over 65 years old, mean personal income, etc. The dataset is available in \cite{smith2002} and 23 out of the 67 vectors contained at least one zero value. 

\begin{figure}[!ht]
\centering
\includegraphics[scale = 0.55]{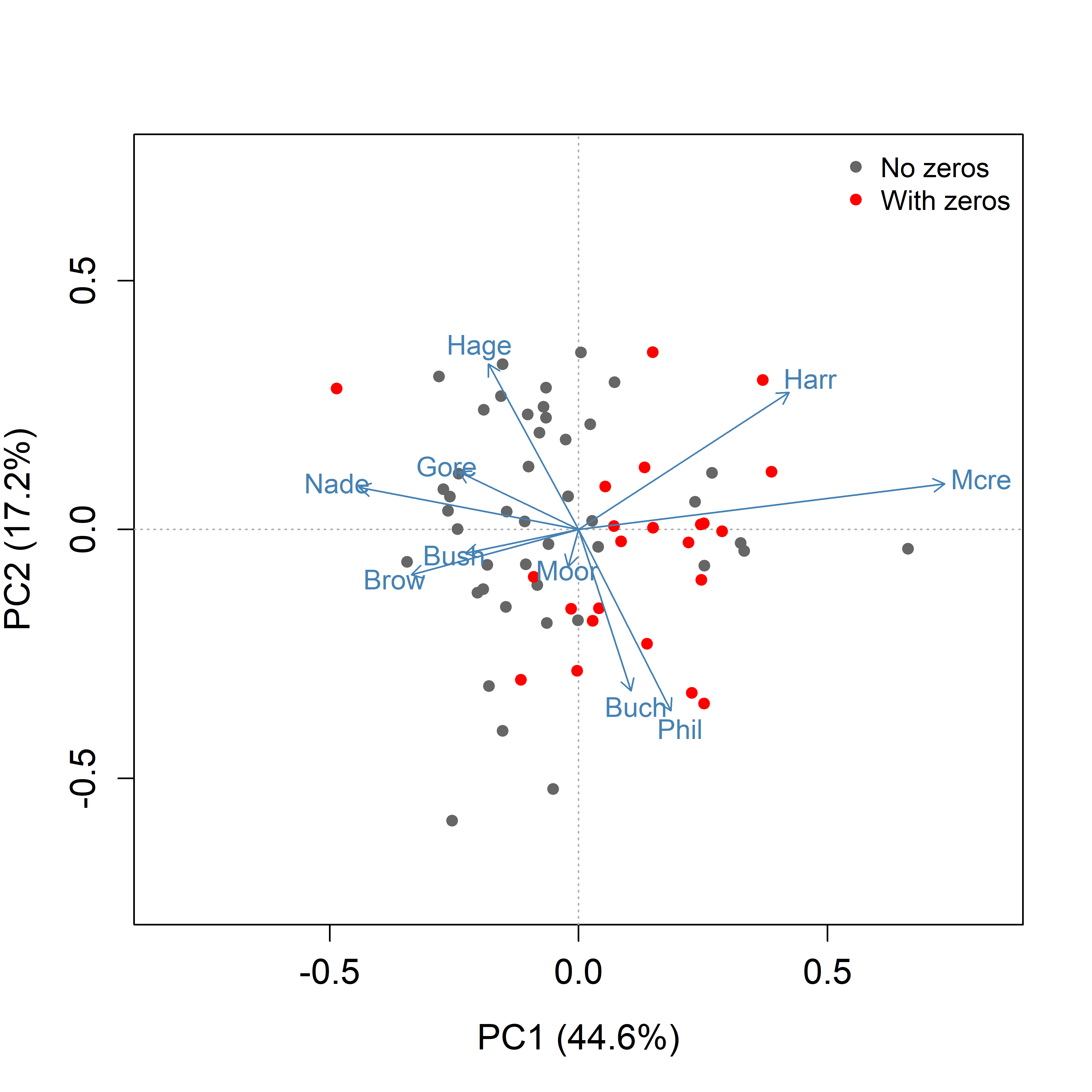} 
\caption{Biplot of the Elections dataset.} 
\label{biplot}
\end{figure}

At first we visualize the data in Figure \ref{biplot} using the biplot. We highlight that the components "Harr", "Hage", "Mcre", "Phil" and "Moor" contain 14, 3, 13, 3, and 3 zero values. These are the components oriented on the right hand side of the biplot, where most of the red points are located.

The computed KLD for the CLNR model was equal to 0.65, while for the ZADR model it was equal to 0.84, again indicating that the CLN model fitted the data better than its competitor. Regarding the computational cost, the CLNR model required 0.03 seconds, whereas ZADR required 2.73 seconds. Since this dataset contains 8 predictor variables we performed variable selection using the FBED algorithm resulting into no variable being discarded.

\begin{figure}[!ht]
\centering
\includegraphics[scale = 0.45, trim = 0 0 0 0]{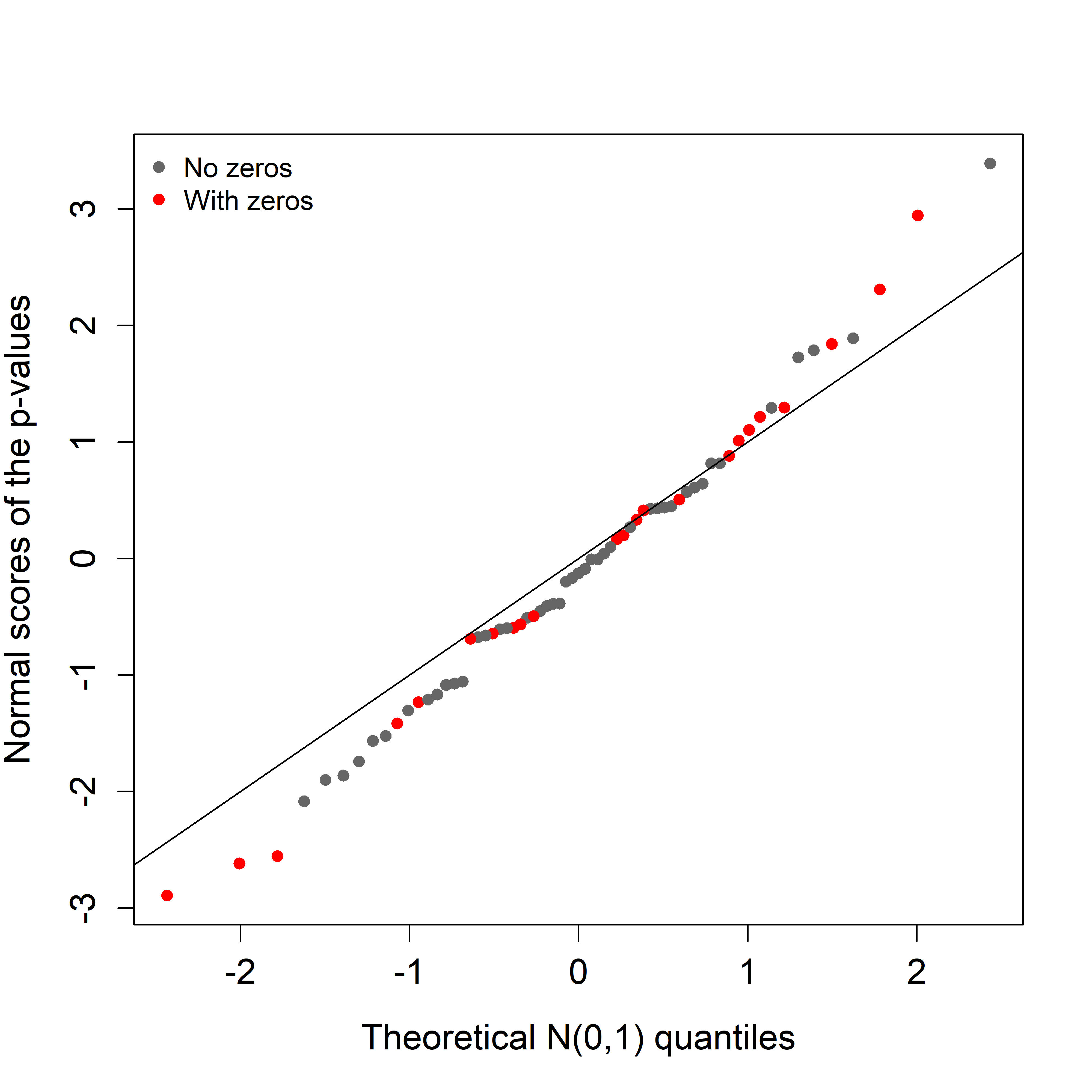} \\
\caption{QQ plot of the residuals of the CLNR model applied to the Elections dataset.} 
\label{resid2}
\end{figure}

Figure \ref{resid2} contains the QQ plot of the CLNR model. Again, the red and grey points, corresponding to compositions with and without zero values present, respectively follow exhibit similar spreads, and the non--parametric energy--based test of distributions provided a $p-value$ equal to 0.912.

\section{Conclusions} \label{sec:concs}
We explored in detail the CLN model initially mentioned by \cite{ait2003} and generalized it to multiple patterns of zero values. We also added covariates in the model and we showed a computationally efficient way to perform MLE at any number of dimensions, by employing the EM algorithm. We compared the CLNR model to its competitor, the ZADR model and showed that the first provides a better fit and it is computationally more efficient. 

The \textit{R} code to implement the CLN and CLNR models is available in the \textit{R} package \textsf{Compositionalcln} \citep{compositionalcln2026}. Our plan is to translate the \textit{R} code to \textit{C++} code to enhance the speed of the EM algorithm. Future work is directed towards exploring the CLNR in more depth, for instance study its residuals, and generalize the model to longitudinal and clustered data. A formal test, or a relaxation in which $\pi$ is linked to $(\bm{\mu}, \bm{\Sigma})$ through shared covariates, is also left for future work. 

\clearpage
\bibliographystyle{apalike}
\bibliography{vivlio}

@article{ait1982,
  title={{The statistical analysis of compositional data}},
  author={Aitchison, J.},
  journal={Journal of the Royal Statistical Society. Series B},
  volume={44},
  number={2},
  pages={139--177},
  year={1982},
  publisher={JSTOR}
}

@book{ait2003,
  title={The statistical analysis of compositional data},
  author={Aitchison, J.},
  year={2003},
  publisher={New Jersey: Reprinted by The Blackburn Press}
}

@article{stewart2011,
  title={Managing the Essential Zeros in Quantitative Fatty Acid Signature Analysis},
  author={Stewart, C. and Field, C.},
  journal={Journal of Agricultural, Biological, and Environmental Statistics},
  volume={16},
  number={1},
  pages={45--69},
  year={2011},
  publisher={Springer}
}

@article{scealy2011,
  title={Regression for compositional data by using distributions defined on the hypersphere},
  author={Scealy, J.L. and Welsh, A.H.},
  journal={Journal of the Royal Statistical Society. Series B},
  volume={73},
  number={3},
  pages={351--375},
  year={2011},
  publisher={Wiley Online Library}
}

@article{leininger2013,
  title={Spatial Regression Modeling for Compositional Data With Many Zeros},
  author={Leininger, Thomas J. and Gelfand, Alan E. and Allen, Jenica M. and Silander Jr, John A.},
  journal={Journal of Agricultural, Biological, and Environmental Statistics},
  volume={18},
  number={3},
  pages={314--334},
  year={2013},
  publisher={Springer}
}

@article{martin2012,
  title={Model-based replacement of rounded zeros in compositional data: Classical and robust approaches},
  author={Mart{\'\i}n-Fern{\'a}ndez, J.A. and Hron, Karel and Templ, Matthias and Filzmoser, Peter and Palarea-Albaladejo, J},
  journal={Computational Statistics \& Data Analysis},
  volume={56},
  number={9},
  pages={2688--2704},
  year={2012},
  publisher={Elsevier}
}

@article{palarea2008,
  title={A modified EM alr-algorithm for replacing rounded zeros in compositional data sets},
  author={Palarea-Albaladejo, J and Mart{\'\i}n-Fern{\'a}ndez, Jose A},
  journal={Computers \& Geosciences},
  volume={34},
  number={8},
  pages={902--917},
  year={2008},
  publisher={Elsevier}
}

@article{zadora2010,
  title={A Two-Level Model for Evidence Evaluation in the Presence of Zeros},
  author={Zadora, Grzegorz and Neocleous, Tereza and Aitken, Colin},
  journal={Journal of Forensic Sciences},
  volume={55},
  number={2},
  pages={371--384},
  year={2010},
  publisher={Wiley Online Library}
}

@article{gueorguieva2008,
  title={Dirichlet component regression and its applications to psychiatric data},
  author={Gueorguieva, Ralitza and Rosenheck, Robert and Zelterman, Daniel},
  journal={Computational statistics \& data analysis},
  volume={52},
  number={12},
  pages={5344--5355},
  year={2008},
  publisher={Elsevier}
}

@article{hijazi2009,
  title={{Modelling compositional data using Dirichlet regression models}},
  author={Hijazi, Rafiq H and Jernigan, Robert W},
  journal={Journal of Applied Probability \& Statistics},
  volume={4},
  number={1},
  pages={77--91},
  year={2009}
}

@Manual{templ2026,
  title={{robCompositions: Robust estimation for compositional data}},
  author={Templ, Matthias and Hron, Karel and Filzmoser, Peter},
  note={R package version  2.6.0},
  url={http://cran.r-project.org/web/packages/robCompositions/},
  year={2026}
}

@article{tsagris2018,
  title={{A Dirichlet regression model for compositional data with zeros}},
  author={Tsagris, Michail and Stewart, Connie},
  journal={Lobachevskii Journal of Mathematics},
  volume={39},
  number={3},
  pages={398--412},
  year={2018},
  publisher={Springer}
}

@article{bear2016,
  title={A logistic normal mixture model for compositional data allowing essential zeros},
  author={Bear, John and Billheimer, Dean},
  journal={Austrian Journal of Statistics},
  volume={45},
  number={4},
  pages={3--23},
  year={2016}
}

@article{smith2002,
  title={{A statistical assessment of Buchanan's vote in Palm Beach county}},
  author="Smith, Richard L",
  journal="Statistical Science",
  volume="17",
  number="4",
  pages="441--457",
  year="2002",
  publisher="JSTOR"
}

@manual{compositions2025,
    title = {{compositions: Compositional Data Analysis}},
    author = {K. Gerald {van den Boogaart} and Raimon Tolosana-Delgado and Matevz Bren},
    year = {2025},
    note = {R package version 2.0-9},
    url = {https://CRAN.R-project.org/package=compositions},
    doi = {10.32614/CRAN.package.compositions},
  }

@article{alenazi2019,
  title={Regression for compositional data with compositional data as predictor variables with or without zero values},
  author={Alenazi, Abdulaziz},
  journal={Journal of Data Science},
  volume={17},
  number={1},
  pages={219--237},
  year={2019}
}

@inproceedings{tsagris2011,
  title={A data-based power transformation for compositional data},
  author={Tsagris, M.T. and Preston, S. and Wood, A.T.A.},
  booktitle={Proceedings of the 4rth Compositional Data Analysis Workshop, Girona, Spain},
  year={2011},
}

@inproceedings{tsagris2015,
  title="A novel, divergence based, regression for compositional data",
  author="Tsagris, M.",
  booktitle="Proceedings of the 28th Panhellenic Statistics Conference, April, Athens, Greece",
  year="2015",
}

@article{chen2016,
  title={A two-part mixed-effects model for analyzing longitudinal microbiome compositional data},
  author={Chen, Eric Z and Li, Hongzhe},
  journal={Bioinformatics},
  volume={32},
  number={17},
  pages={2611--2617},
  year={2016},
  publisher={Oxford University Press}
}

@article{rubin1976,
  author  = {Rubin, Donald B.},
  title   = {Inference and missing data},
  journal = {Biometrika},
  year    = {1976},
  volume  = {63},
  number  = {3},
  pages   = {581--592}
}

@manual{compositionalcln2026,
    title = {{Compositionalcln: Modelling Compositional Data with Zero Values}},
    author = {Michail Tsagris},
    year = {2026},
    note = {R package version 1.0},
  }

@article{louis1982,
  author  = {Louis, Thomas A.},
  title   = {{Finding the Observed Information Matrix When Using the EM Algorithm}},
  journal = {Journal of the Royal Statistical Society: Series B (Methodological)},
  year    = {1982},
  volume  = {44},
  number  = {2},
  pages   = {226--233}
}

@article{tsagris2026,
  title={The $\alpha$--regression for compositional data: a unified framework for standard, temporal and spatial regression models including compositional predictors},
  author={Tsagris, Michail and Alharbi, Nader and Alenazi, Abdulaziz and Pantazis, Yannis},
  journal={arXiv preprint arXiv:2510.12663},
  year={2026}
}

@article{tibshirani1996,
  title={{Regression Shrinkage and Selection via the Lasso}},
  author={Tibshirani, Robert},
  journal={Journal of the Royal Statistical Society Series B: Statistical Methodology},
  volume={58},
  number={1},
  pages={267--288},
  year={1996},
  publisher={Oxford University Press}
}

@article{borboudakis2019,
  title={{Forward-Backward Selection with Early Dropping}},
  author={Borboudakis, Giorgos and Tsamardinos, Ioannis},
  journal={Journal of Machine Learning Research},
  volume={20},
  number={8},
  pages={1--39},
  year={2019}
}

@article{li2018,
  title={Conditional regression based on a multivariate zero-inflated logistic-normal model for microbiome relative abundance data},
  author={Li, Zhigang and Lee, Katherine and Karagas, Margaret R and Madan, Juliette C and Hoen, Anne G and O’malley, A James and Li, Hongzhe},
  journal={Statistics in biosciences},
  volume={10},
  number={3},
  pages={587--608},
  year={2018},
  publisher={Springer}
}

@article{ait2002,
  title={Biplots of compositional data},
  author={Aitchison, John and Greenacre, Michael},
  journal={Journal of the Royal Statistical Society Series C: Applied Statistics},
  volume={51},
  number={4},
  pages={375--392},
  year={2002},
  publisher={Oxford University Press}
}

@article{ait1983,
  title={Principal component analysis of compositional data},
  author={Aitchison, John},
  journal={Biometrika},
  volume={70},
  number={1},
  pages={57--65},
  year={1983},
  publisher={Oxford University Press}
}

@article{tsagris2025b,
  title={{Simplicial Clustering Using the $\alpha$-Transformation}},
  author={Tsagris, Michail and Adam, Christos and Kontemeniotis, Nikolaos},
  journal={Lobachevskii Journal of Mathematics},
  volume={46},
  number={12},
  pages={6471--6482},
  year={2025},
  publisher={Springer}
}

@article{sevincc2026,
  title={Energy based equality of distributions testing for compositional data},
  author={Sevin{\c{c}}, Volkan and Tsagris, Michail},
  journal={Communications in Statistics-Simulation and Computation},
  pages={1--15},
  year={2026},
  publisher={Taylor \& Francis}
}

@article{james1954,
 title={Tests of linear hypotheses in univariate and multivariate analysis when the ratios of the population variances are unknown},
  author={James, G.S.},
  journal={Biometrika},
  volume={41},
  number={1/2},
  pages={19--43},
  year={1954},
  publisher={JSTOR}
}

\end{document}